\documentclass[11pt]{emulateapj}

\usepackage{amsmath}
\usepackage{amssymb}
\usepackage{color}
\usepackage{enumitem}
\usepackage{graphicx}
\usepackage{mathptmx}
\usepackage{newtxmath}
\usepackage{newtxtext}
\usepackage{natbib}

\setlist[itemize]{leftmargin=*}
\setlist[enumerate]{leftmargin=*}

\begin{document}
\newcommand{\av}[1]{\left<#1\right>}
\renewcommand{\vec}[1]{\mathbf{#1}}
\newcommand{\hvec}[1]{\hat{\mathbf{#1}}}
\newcommand{\tens}[1]{{\mathsf{#1}}}
\newcommand{\pd}[2]{\frac{\partial #1}{\partial #2} }
\newcommand{\HALF}{\frac{1}{2}}
\newcommand{\indx}{{\vec{\mathrm i}}}
\newcommand{\THREEHALF}{\frac{3}{2}}
\newcommand{\DS}{\displaystyle}
\newcommand{\Fluid}{\color{blue}Fluid\color{black}}
\newcommand{\Particles}{\color{red}Particles\color{black}}
\newcommand{\quotes}[1]{``#1''}
\newcommand{\C}{{\mathbb C}}
\newcommand{\D}{{\textsc{\scriptsize D}}}
\newcommand{\F}{{\cal F}}
\newcommand{\T}{{\cal T}}
\newcommand{\RED}{\color{red}}
\newcommand{\BLACK}{\color{black}}
\newcommand{\BLUE}{\color{black}}
\newcommand{\GREEN}{\color{green}}

\title{A Particle Module for the PLUTO Code: III - Dust}


\author{A. Mignone}
\affil{Dipartimento di Fisica, Universit\'a di Torino
       via Pietro Giuria 1 (10125) Torino, Italy}

\author{M. Flock}
\affil{Max Planck Institute f\"ur Astronomy (MPIA), K\"onigstuhl 17, 69117 Heidelberg, Germany}

\and

\author{B. Vaidya}
\affil{Discipline of Astronomy, Astrophysics and Space Engineering, Indian Institute of Technology Indore, Khandwa Road, Simrol, 453552 , India }

\altaffiltext{1}{AAS Journals Data Scientist}

\begin{abstract}
The implementation of a new particle module describing the physics of dust grains coupled to the gas via drag forces is the subject of this work.
The proposed particle-gas hybrid scheme has been designed to work in Cartesian as well as in cylindrical and spherical geometries.
The numerical method relies on a Godunov-type second-order scheme for the fluid and an exponential midpoint rule for dust particles which overcomes the stiffness introduced by the linear coupling term.
Besides being time-reversible and globally second-order accurate in time, the exponential integrator provides energy errors which are always bounded and it remains stable in the limit of arbitrarily small particle stopping times yielding the correct asymptotic solution.
Such properties make this method preferable to the more widely used semi-implicit or fully implicit schemes at a very modest increase in computational cost.
Coupling between particles and grid quantities is achieved through particle deposition and field-weighting techniques borrowed from Particle-In-Cell simulation methods.
In this respect, we derive new weight factors in curvilinear coordinates that are more accurate than traditional volume- or area-weighting.

A comprehensive suite of numerical benchmarks is presented to assess the accuracy and robustness of the algorithm in Cartesian, cylindrical and spherical coordinates.
Particular attention is devoted to the streaming instability which is analyzed in both local and global disk models.
The module is part of the PLUTO code for astrophysical gas-dynamics and it is mainly intended for the numerical modeling of protoplanetary disks in which solid and gas interact via aerodynamic drag.  
\end{abstract}


\section{Introduction}
\label{sec:Introduction}
%
%
%
%
The dynamics of gas and dust is an essential ingredient in the star and planet formation area.
Especially with the recent advancements in scattered light and mm continuum observations of dust grains in protoplanetary disks \citep{Avenhaus.2018,Andrews.2018}, we have to understand the dust characteristics of distribution, size and internal properties. 
The motion of dust grains embedded in a protoplanetary disk depends crucially on their grain size. 
A convenient parameter describing this coupling between dust and gas components is the stopping time $\tau_s$ \citep{Whipple.1972,Weidenschilling.1977} or often expressed as the Stokes number ${\rm St} = \tau_s \Omega_K$
\BLUE, where $\Omega_K$ is the Keplerian angular velocity\BLACK. 
Small dust grains below one micron are coupled to the motion of the gas due to the strong drag force by the gas ${\rm St} \ll 1$. Solid bodies larger than 10 meter become practically decoupled from the gas motion ${\rm St} \gg 1$. 
One of the most crucial problem in the formation of planets from small grains is the so called radial-drift barrier.
As grains grow in size and their Stokes number approaches unity they quickly drift to the star: for the minimum-mass solar nebula model \citep{Hayashi1981}, for instance, this process takes place on a timescale of approximately $100$ yr at $\sim 1$ AU for a meter-sized solid body \citep{Weidenschilling.1977}.

The physics of dust grains plays also an important role in shaping the morphology of circumstellar environments around bow shocks for fast moving red supergiant stars \citep[see, e.g.,][and references therein]{vanMarle_etal.2011, Meyer_etal.2014, Thun_etal.2016}.
As dust grains are a primary source of infrared radiation, their imprint on the surroundings can be investigated by means of numerical simulations, allowing infrared emission maps to be constructed \citep[see, e.g.][]{Meyer_etal.2017} and ultimately validated against observational evidence.

The need for sophisticated numerical tools to solve the coupled dust and gas equations becomes therefore crucial in the attempt of bridging the gap between theoretical models and observational data.
In the case of protoplanetary disk, this became again evident with the recent observations of the protoplanetary disk HL Tau \citep{ALMA2015} with the Atacama Large Millimeter/submillimeter Array (ALMA).
Best fit models of HL Tau predict grains of millimeter and submillimeter size responsible for the emission \citep{Pinte.2016, Carrasco.2016, Liu.2017}.
Such grains sizes, located at resolvable disk radii above 10 AU, have Stokes numbers close to unity and coupled dust and gas models are necessary to explain the observed structures in this disk. 

In recent years, a variety of numerical methods for the solution of the composite (gas+dust) system have been proposed.
While the gas component is treated using conventional grid-based schemes, dust can be either modeled as a pressureless fluid \citep[see, e.g.,][and references therein]{JK.2005, Miniati.2010, Meheut_etal.2012, Kowalik_etal.2013, Laibe_Price.2014, Porth_etal.2014, Llambay_etal.2019} or as a system of super-particles as in \cite{YJ.2007} (YJ07 hereafter), \cite{Balsara_etal.2009}, \cite{Bai_Stone.2010} (BS10 hereafter), \cite{Zhu_etal.2014}, \cite{Yang_Johansen.2016}.
Compared to the former, particle (or Lagrangian) methods have the advantage to follow the dynamical and thermal history of individual grains in the disk.
Besides, they allow to resolve crossing trajectories between grains in a single computational zone and their individual motion which becomes important once the grains reach Stokes numbers close to unity.

Dynamic coupling between dust and fluid components takes place through aerodynamic drag forces causing the two components to exchange momentum through mutual feedback terms.
Despite its simplicity (feedback is modeled as a linear term in the relative velocity), the drag acceleration can become stiff for tightly coupled grains ($\tau_s \ll \Delta t$), thus forcing an explicit scheme to abnormally small time stepping.
To overcome this limitation, explicit integration schemes of dust grains are superseded by either semi- or fully-implicit \citep[see BS10 or][]{Zhu_etal.2014} or semi-analytical approaches based on operator splitting techniques \cite[see][]{Yang_Johansen.2016}.
Although the majority of the aforementioned studies has succeeded in presenting state-of-the-art hybrid numerical schemes for modeling fluid and dust particles with mutual feedback in Cartesian coordinates, the extension of dust particle-gas hybrid schemes to cylindrical and/or spherical geometries has so far been proposed in the context of test particles only, neglecting the back-reaction of the dust onto the gas \citep{Zhu_etal.2014, Stoll_Kley.2016, Picogna_etal.2018}.

With the present work we intend to overcome this limitation by presenting a new particle-gas hybrid scheme suitable for the solution of the dust-fluid coupled equations in Cartesian, cylindrical and spherical geometries.
Our formulation introduces two innovative major aspects, i) the employment of an exponential integrator for the numerical solution of the particles equations of motions and ii) the inclusion of back-reaction terms from particles onto the the gas through improved particle-in-cell (PIC) field-weighting techniques in curvilinear geometry.
As we shall show, the exponential midpoint rule \citep{HO_2010} has a number of desirable properties: it possesses second-order accuracy, time-reversibility, bounded energy errors during the integration while remaining asymptotically stable for any particle stopping time.
The numerical model, implemented in the PLUTO code \citep{PLUTO.2007}, follows up on preceding hybrid particle-fluid schemes developed in the framework of the MHD-PIC equations \cite[][Paper I henceforth]{Mignone_etal.2018} and to address non-thermal emission from highly energetic particles embedded in a large-scale magnetohydrodynamic (MHD) flows \citep{Vaidya_etal.2018}.

The paper is organized as follows.
In section \ref{sec:equations} we review the fundamental basic equations in different coordinate systems (Cartesian, cylindrical and spherical).
The numerical method is described in section \ref{sec:implementation} and numerical benchmarks are presented in section \ref{sec:tests}.
Conclusion are finally drawn in section \ref{sec:summary}.

\section{Relevant Equations}
\label{sec:equations}
%

We now present the detailed form of the fluid-dust equations.
In what follows we adopt the same notation used in paper I, with $\vec{v}_g,\, \rho,\, p$ and $E_g$ denoting, respectively, the velocity, density, thermal pressure and total energy of the gas:
\begin{align}
  \pd{\rho}{t} & + \nabla\cdot(\rho\vec{v}_g) = 0 \,,
  \label{eq:continuity} \\ \noalign{\smallskip}
  \pd{(\rho\vec{v}_g)}{t} &+ \nabla\cdot
  \left(\rho\vec{v}_g\vec{v}_g\right) + \nabla p = \rho\vec{a}_g 
   + \av{\vec{f}_\D}  \,,
  \label{eq:momentum} \\ \noalign{\smallskip}
  \pd{E_g}{t} &+ \nabla\cdot\left[\left(E_g + p\right)\vec{v}_g\right] 
  = \vec{v}_\D\cdot\av{\vec{f}_\D} \,.
  \label{eq:energy}
\end{align}
%

The first term on the right hand side of Eq. (\ref{eq:momentum}), $\rho \vec{a}_g$, stands for external forces (i.e. gravity, Coriolis, etc..) acting on the fluid.
Its explicit form depends on the particular equation model adopted.
The second term in Eq. (\ref{eq:momentum}) represents the average cumulative drag force accounting for feedback from dust particles to the gas.
In the Epstein regime, the size of dust grains $a$ is much smaller than the mean free path of gas molecule $\lambda$ \citep[$a < 9\lambda/4$,][]{Weidenschilling.1977},
and the drag force is proportional to the relative velocity between the two components,
\begin{equation}\label{eq:gas_fdrag}
  \vec{f}_{\D} = -\rho_\D\frac{\vec{v}_g - \vec{v}_\D}{\tau_s} \,,
\end{equation}
where $\rho_\D$, $\vec{v}_\D$ and $\tau_s$ are, respectively the dust density, velocity and stopping (friction) time.
For a typical protoplanetary disk, the mean free path of gas molecules is larger than $1$ cm for regions outside 1 AU \citep{NSH.1986}.
In this regime the stopping times becomes 
\begin{equation}
  \tau_s \approx \frac{\rho_{\bullet}a}{\rho(R) v_t} \,,
\end{equation}
where $\rho_{\bullet}$ is the solid material density and $v_t= \sqrt{8/\pi} c_s$ is the mean thermal velocity with the sound speed $c_s$ \citep{Takeuchi.2001}.

The source term in the (gas) total energy equation accounts for the work done by the drag force and the frictional heating due to dust \citep[see, e.g.,][]{Marble.1970, Pelanti_LeVeque.2006, Laibe_Price.2014},
\begin{equation}
  \vec{v}_\D\cdot\av{\vec{f}}_\D =  \vec{v}_g\cdot\av{\vec{f}}_\D
                                  + \rho_\D\frac{(\vec{v}_g-\vec{v}_\D)^2}{\tau} \,.
\end{equation}
The previous expression follows directly from total energy conservation of the gas+dust mixture.
In the present work, however, we shall neglect the energy equation and restrict the attention to a locally isothermal gas for which density and pressure are related by the  equation of state
\begin{equation}\label{eq:isothermal_eos}
  p = \rho c_s^2 \,,
\end{equation}
where $c_s=c_s(\vec{x})$ is the (local) isothermal speed of sound.
The isothermal approximation holds when the thermal relaxation timescale is much shorter than the dynamical timescale.
For typical protoplanetary disks this assumption is verified when the plasma becomes optical thin to thermal radiation, which is typically the case for the outer disk regions.

The dust component is modeled as an ensemble of macro-particles governed by the system of ordinary differential equations (ODEs)
\begin{equation}\label{eq:dust_ode}
  \left\{\begin{array}{lcl}
    \DS \frac{d\vec{x}_p}{dt}  &=& \vec{v}_p  \,,
    \\ \noalign{\smallskip}
    \DS \frac{d\vec{v}_p}{dt} &=& \DS  \vec{a}_p
                                - \frac{\vec{v}_p-\vec{v}_g}{\tau_{s,p}} \,,
  \end{array}\right.
\end{equation}
where, keeping the same notation as in paper I, $\vec{x}_p$ and $\vec{v}_p$ are used to denote the dust particle position and velocity vectors, $\vec{a}_p$ is the acceleration vector that includes external forces (without the gas-particle drag term).

We now consider Eqns (\ref{eq:continuity})-(\ref{eq:energy}) together with (\ref{eq:dust_ode}) in the shearingbox approximation, widely used for local disk models, and in cylindrical and spherical coordinates which are best suited to describe a global scenario. 
\subsection{Shearing Box Equations Model}
%

The shearing-box approximation \citep{HawGamBal.1995} provides a local model of a differentially rotating disk that closes on a small rectangular patch in a frame co-rotating with the disk at some fiducial radius.
In this non-inertial frame a local Cartesian system of coordinates is introduced with $x,y$ and $z$ pointing in the radial, azimuthal and vertical directions, respectively.
The background orbital motion is described by a linear velocity shear $\vec{w} = -q\Omega x\hvec{e}_y$, where $\Omega$ is the local constant angular velocity and $q$ is a local measure of the differential rotation ($q = 3/2$ for a Keplerian profile).
Boundary conditions are shear-periodic, meaning that a gas element on either side of the computational domain shifts with respect to the one on the opposite side by a finite amount given $|\vec{w}| t$. 

In the standard shearingbox equations, the acceleration term $\vec{a}_g$ accounts for the Coriolis term, centrifugal and gravitational forces:
\begin{align}
   \label{eq:standard_SB_g}
    \vec{a}_g &=\DS  2\vec{v}_g\times\vec{\Omega}
                    + \Omega^2\left(2qx\hvec{e}_x - z\hvec{e}_z\right)
                    + 2\Omega \eta v_K \hvec{e}_x \,, \\ 
  \label{eq:standard_SB_p}
    \vec{a}_p &=\DS  2\vec{v}_p\times\vec{\Omega}
                    + \Omega^2\left(2qx\hvec{e}_x - z\hvec{e}_z\right) \,.
\end{align}
where $\vec{\Omega} = \Omega\hvec{e}_z$.
The parameter $\eta$ describes the disk global pressure gradient through a dimensionless measure of sub-Keplerian rotation,
\begin{equation}\label{eq:eta}
  \eta \equiv - \frac{c_s^2}{2\rho v_K^2}\pd{\rho}{\log r}
       \approx \left(\frac{c_s}{v_K}\right)^2 \,,
\end{equation}
where $c_s$ is the sound speed while $v_K$ is the local Keplerian speed \citep[see, e.g.][YJ07 and BS10]{YG.2005}.
Stated differently, it measures the amount by which the gas azimuthal velocity is reduced from the Keplerian value due to the radial pressure gradient.
The quantity $\eta v_K$ is here added to the gas and it produces a constant acceleration term pointing outward.

An orbital advection scheme can be used to subtract the linear shear contribution from the total velocity so that the hydro or MHD equations can be evolved only in the residual thus giving a substantial speedup of the algorithm.
This entails a redefinition of the total velocity $\vec{v} = \vec{v}' - q\Omega x\hvec{e}_y$ where the first term corresponds to the residual and the second term represents the mean shear.
The implementation of the FARGO scheme for orbital advection in the PLUTO code is thoroughly described in the work by \cite{Mignone_FARGO.2012} and it can be employed for local or global disk simulations.
When used in conjunction with the FARGO orbital scheme, Eq. (\ref{eq:standard_SB_g}) and (\ref{eq:standard_SB_p}) are modified into
\begin{align}
  \label{eq:FARGO_SB_g}
  \vec{a}'_g &=    2\vec{v}'_g\times\vec{\Omega}
               + qv'_{g,x}\Omega\hvec{e}_y
               - \Omega^2z\hvec{e}_z
               + 2\Omega \eta v_K \hvec{e}_x \,, \\ 
  \label{eq:FARGO_SB_p}
  \vec{a}'_p &=    2\vec{v}'_p\times\vec{\Omega}
               + qv'_{p,x}\Omega\hvec{e}_y 
               - \Omega^2z\hvec{e}_z  \,.
\end{align}
where the velocity are now the residual ones.
Unless otherwise stated, we use $q = 3/2$ typical for a Keplerian profile.

\subsection{Equations in Cylindrical and Spherical Coordinates}
\label{eq:Cylindrical_and_Spherical}
%

While the explicit expressions for the gas-dynamical and MHD equations in cylindrical an spherical coordinates can be found in \cite{Mignone_FARGO.2012}, in what follows we give the explicit expressions of the particle equation of motion (\ref{eq:dust_ode}) in these coordinate systems.

In cylindrical coordinates, Eq. (\ref{eq:dust_ode}) is solved in terms of the particle coordinates $(R,\phi,z)$, radial and vertical velocities $v_R$ and $v_z$ and the angular momentum $l_\phi \equiv Rv_{\phi}$ (we drop the subscript $p$ to avoid cluttered notations):
\begin{equation}\label{eq:dust_cylindrical}
  \begin{array}{lcl}
    \DS \frac{dR}{dt}     & = & \DS v_{R}  \,,              \\ \noalign{\smallskip}
    \DS \frac{d\phi}{dt}  & = & \DS \frac{l_\phi}{R^2}  \,, \\ \noalign{\smallskip}
    \DS \frac{dz}{dt}     & = & \DS v_{z}  \,,       \\ \noalign{\smallskip}

    \DS \frac{dv_{R}}{dt} & = & \DS    \frac{l_\phi^2}{R^3}
                                       + g_R
                                       - \frac{v_R - v_{g,R}}{\tau_s}\,,
    \\ \noalign{\smallskip}
    \DS \frac{dl_\phi}{dt}& = & \DS      Rg_\phi
                                       - \frac{l_\phi - l_{g,\phi}}{\tau_s}\,,
\\ \noalign{\smallskip}
    \DS \frac{dv_{z}}{dt} & = & \DS    g_z
                                       - \frac{v_{z} - v_{g,z}}{\tau_s} \,.
  \end{array}
\end{equation}

%

Likewise, in spherical coordinates, we solve for the particle coordinates $(r,\theta,\phi$) and $v_r$, $l_\theta = rv_\theta$, $l_\phi = rv_\phi\sin\theta$ (angular momentum):
\begin{equation}\label{eq:dust_spherical}
  \begin{array}{lcl}
  \DS \frac{dr}{dt}       & = & \DS v_{r} \,,
  \\ \noalign{\smallskip}
  \DS \frac{d\theta}{dt}  & = & \DS \frac{l_\theta}{r^2}\,,
  \\ \noalign{\smallskip}
  \DS \frac{d\phi}{dt}    & = & \DS \frac{l_\phi}{r^2\sin^2\theta}\,,
  \\ \noalign{\smallskip}
  \DS \frac{d v_r}{d t} &=& \DS  \frac{l_\theta^2}{r^3}
                               + \frac{l_\phi^2}{r^3\sin^2\theta} + g_r
                               - \frac{v_r - v_{g, r}}{\tau_s}    \,,
  \\  \noalign{\smallskip}
  \DS \frac {dl_\theta}{dt} &=& \DS  \frac{l_\phi^2 \cos\theta}{r^2\sin^3\theta}
                                    + r g_\theta
                                    - \frac{l_\theta - v_{g, \theta}r}{\tau_s}\,,
   \\  \noalign{\smallskip}
       \DS \frac{d l_{\phi}}{d t}\, \, &=& \DS r \sin \theta g_\phi
                              - \frac{l_\phi - l_{g, \phi}}{\tau_s} \,.
  \end{array}
\end{equation}
%
%
In the previous expressions, $l_{g,\phi}$ is used to denote the gas angular momentum.

Orbital advection can be incorporated by decomposing the particle velocity into average azimuthal contribution and a residual term, $v_\phi = v'_\phi + \av{\Omega}R$  where $\av{\Omega}$ is the mean gas angular velocity along the azimuthal direction \citep{Mignone_FARGO.2012}.
However, due to lengthy reason, we shall present this variant in a companion paper.

\section{Numerical Approach}
\label{sec:implementation}
%
%
%

The gas-dynamical Equations (\ref{eq:continuity})-(\ref{eq:energy}) together with the dust particle equations of motion (\ref{eq:dust_ode}) are solved within the same framework developed for the PLUTO code and thus using an approach similar to that outlined in paper I.
Fluid quantities are advanced in time using a conservative scheme already available in the code.
The hydro solvers are coupled to the particle integrator and have been properly modified to account for the dust back-reaction.
Particles are stored in memory using a doubly linked list, which consists of sequentially linked node structures (see paper I).
Each particle is defined by its spatial coordinates $\vec{x}_p$, velocity $\vec{v}_p$, stopping time $\tau_{s,p}$ and mass $m_p$.
Particles can be pushed using two different integrators which we describe in section \ref{sec:particle_integrator}.

\subsection{Hydro Integrators}
\label{sec:hydro_integrator}
%

The gas-dynamical equations are discretized on a computational grid with cell indices $\indx\equiv(i,j,k)$ and advanced in time using a second-order method typically consisting of a predictor-corrector scheme already available in the code such as the Corner Transport Upwind \citep[CTU,][]{Colella.1990, PLUTO.2012} or the Runge-Kutta method \citep[RK2][]{Gottlieb_Shu.1998}.
While the use of the former is recommended for Cartesian geometries, the latter can be more easily adapted to cylindrical or spherical coordinates.

In the case of RK2, for instance, the predictor step is achieved through
\begin{equation}\label{eq:gas_predictor}
  U_\indx^{*} = U^n_\indx + \Delta t^n\Big(
               {\cal L}^n_\indx + S^n_{\D,\indx}\Big) \,,
\end{equation}
where conservative quantities are denoted by the array $U=\{\rho,\, \rho\vec{v}_g\}$ (magnetic fields can be easily added to our formulation), ${\cal L}^n_\indx$ accounts for the conservative flux-difference operator as well as additional source terms not tied to particles while $S^n_{\D,\indx} = (0,\, \av{\vec{f}_\D})^n_\indx$ is a numerical approximation to the gas-dust coupling term (\ref{eq:gas_fdrag}).
The flux-difference operator has the form
\begin{equation}\label{eq:Lop}
  {\cal L}^n_\indx =  -\frac{1}{\Delta V_\indx}\sum_d
              \Big[ \left(A\F\right)_{\vec{i}+\HALF\hvec{e}_d}
                   -\left(A\F\right)_{\vec{i}-\HALF\hvec{e}_d}\Big]
              + S^n_{g,\indx} \,,
\end{equation}
where $\F$ is the numerical flux computed at cell interfaces by means of a Riemann solver, \BLUE $A$ is the interace area, $\Delta V_{\indx}$ is the cell volume \BLACK and $S^n_{g,\indx}$ accounts for external forces (e.g. Eq. \ref{eq:standard_SB_g} or \ref{eq:FARGO_SB_g} for the shearingbox model) as well as geometrical source terms, if any.
Explicit expressions may be found, e.g., in \cite{Mignone_FARGO.2012}.
\BLUE
The left and right input states to the Riemann solver are obtained by standard reconstruction techniques see, e.g., \cite{PLUTO.2012} for the Cartesian coordinates case or \cite{Mignone.2014} for cylindrical and spherical grids.
\BLACK

The back-reaction coupling term depends on the relative velocity between the two species and it is obtained by first computing the drag acceleration at the particle position,
\begin{equation}\label{eq:Delta_vp}
  \Delta\vec{v}^n_{p} = \left(\sum_{\indx}W(\vec{x}^n_p-\vec{x}_{\indx})
                        \vec{v}^n_{g,\indx}\right) - \vec{v}^n_p \,,
\end{equation}
and then by distributing them back on the grid,
\begin{equation}\label{eq:f_D}
  \av{\vec{f}^n_{\D}}_\indx = -\frac{1}{\Delta V_\indx}\sum_p \frac{m_p}{\tau_{s,p}}
                          W(\vec{x}^n_p - \vec{x}_\indx)\Delta \vec{v}^n_{p} \,.
\end{equation}
This ensures momentum conservation and helps in reducing spurious acceleration effects \citep[][and YJ07]{JY.2007}.
In Eq. (\ref{eq:Delta_vp}) the summation extends to all grid zones and the weights $W(\vec{x}_p-\vec{x}_i)$ are functions of the relative distance between particle $p$ and the grid zone $\indx$.
Weights function are discussed in section \ref{sec:weights}.
As the particle support extends over nearest neighbor zones, the summation is restricted just to $9$ zones in two dimensions and $27$ in three dimensions.
The same argument applies to Eq. (\ref{eq:f_D}).

After the predictor stage, particles can be pushed for a full step using the fluid velocity at the half time level, $(\vec{x}_p,\, \vec{v}_p)^n \to (\vec{x}_p,\, \vec{v}_p)^{n+1}$.
This is detailed in Section \ref{sec:particle_integrator}.

The fluid variables are then updated in the corrector step, e.g., 
\BLUE
\begin{equation}
  U^{n+1}_\indx = U^n_\indx + \frac{\Delta t^n}{2}
                 \Big(  {\cal L}_\indx^{n} 
                      + {\cal L}_\indx^{*}\Big)
                  + \Delta t^nS_{\D,\indx}^{n+\HALF}\,,
\end{equation}
\BLACK
where $S^{n+\HALF}_{\D,\indx}=(0,\av{\vec{f}_\D})^{n+\HALF}_\indx$ can now be more conveniently expressed in terms of the momentum variation of individual particles, after deducting non-frictional forces:
\begin{equation}\label{eq:dust_feedback_final}
  \vec{f}_{\D,\indx}^{n+\HALF}
    = -\frac{1}{\Delta V_\indx}\sum_p m_pW(\vec{x}^{n+\HALF}_p - \vec{x}_\indx)
             \left(\Delta\vec{v}_p^{n+\HALF} - \Delta t^n \vec{a}_p^{n+\HALF}
             \right) \,,
\end{equation}
where $\Delta\vec{v}^{n+\HALF}_p = \vec{v}_p^{n+1} - \vec{v}^n_p$ is the velocity change.
\BLUE
This guarantees total momentum conservation for the composite (gas+dust) system.
The derviation of the CTU scheme follows the same line and it is reported in the Appendix \ref{app:CTU} (see also paper I).

We point out that stiffness may arise during the gas predictor step update in presence of strong local concentrations \citep{Yang_Johansen.2016}, leading the coupling term to diverge in Eq. (\ref{eq:f_D}).
This issue has also been addresses by BS10, who suggested to modify the particle stopping time appearing in the denominator of Eq. (\ref{eq:f_D}) by with $\tau_{s,p} \leftarrow \max(\tau_{s,p}, \Delta t)$.
Here adopt the same strategy albeit, in forthcoming works, we will explore alternative, stiff-accurate time-marching schemes for the fluid component.
\BLACK

\subsection{Particle Integrators}
\label{sec:particle_integrator}
%

In this section we describe and compare two particle integrators, namely: the widely used semi-implicit method and a new, powerful method which relies on exponential integrators.
As we shall see, the latter approach is time-reversible, has the correct asymptotic solution and, unlike the semi-implicit method, it remains stable for any stopping time.

In order to avoid cluttered notations we drop, in the following subsections, the subscript $p$ when referring to particles quantities.

\subsubsection{The Semi-Implicit Method}
\label{sec:semi_implicit}

The semi-implicit integrator already presented by BS10 is a second-order position-Verlet scheme consisting of a drift-kick-drift steps.
Omitting the $p$ subscript,
\begin{equation}\label{eq:semi-implicit}
  \begin{array}{lcl}
    \vec{x}^{n+\HALF} &=&\DS \vec{x}^n + \frac{\Delta t}{2}\vec{v}^{n}
    \\ \noalign{\smallskip}
    \vec{v}^{n+1} &=& \DS \vec{v}^n
                   + \Delta t \vec{a}_{\rm tot}\Big(\vec{x}^{n+\HALF}, \,
                                          \vec{v}^{n+\HALF}, \,
                                          \vec{v}_g^{n+\HALF}\Big)
    \\ \noalign{\smallskip}
    \vec{x}^{n+1} &=&\DS \vec{x}^{n+\HALF} + \frac{\Delta t}{2}\vec{v}^{n+1}
  \end{array}
\end{equation}
where $\vec{a}_{\rm tot}$ is the total acceleration (external plus drag term), $\vec{v}^{n+\HALF}$ is a shorthand notation for $(\vec{v}^{n} + \vec{v}^{n+1})/2$.
The velocity update step is implicit, although a simple Taylor expansion can be used to linearize the acceleration vector,
\begin{equation}\label{eq:a_linear}
  \vec{a}^{n+\HALF}_{\rm tot} \approx \vec{a}^n_{\rm tot}
    + \left(\pd{\vec{a}_{\rm tot}}{\vec{v}}\right)^n
      \frac{\vec{v}^{n+1} - \vec{v}^n}{2} \,,
\end{equation}
where $\partial\vec{a}_{\rm tot}/\partial\vec{v}$ denotes the Jacobian of the acceleration with respect to the particle velocity compute at the base time level.
Inserting the previous expression into the second of Eqns. (\ref{eq:semi-implicit}) leads to the following velocity update
\begin{equation}
  \vec{v}^{n+1} = \vec{v}^n + \tens{\Lambda}^{-1}\Delta t\vec{a}^n_{\rm tot} \,,
\end{equation}
where
\begin{equation}\label{eq:Lambda}
  \mathsf{\Lambda} = \tens{I} - \frac{\Delta t}{2}\pd{\vec{a}_{\rm tot}}{\vec{v}} \,.
\end{equation}
For a simple drag term in absence of external forces ($\vec{a}_p = 0$ in Eq. \ref{eq:dust_ode}), $\Lambda$ is a the diagonal matrix $\Lambda = {\rm diag}(1+b)$ where $b=\Delta t/2\tau_s$.

In the standard Shearing Box model (Eq. \ref{eq:standard_SB_g}), straightforward differentiation leads to 
\begin{equation}
  \tens{\Lambda} = \left(\begin{array}{ccc}
     1 + b           & -\Omega\Delta t  & 0 \\ \noalign{\smallskip}
     \Omega\Delta t  & 1 + b            & 0  \\ \noalign{\smallskip}
               0     &       0          & 1 + b
  \end{array}\right) \,.
\end{equation}
When used in conjunction with the orbital scheme, the corresponding $\Lambda$ matrix is modified into
\begin{equation}
  \tens{\Lambda} = \left(\begin{array}{ccc}
     1 + b & -\Omega\Delta t       & 0 \\ \noalign{\smallskip}
     (1-q/2)\Omega\Delta t  & 1 + b& 0  \\ \noalign{\smallskip}
               0            &       0 & 1 + b
  \end{array}\right) \,.
\end{equation} 

As outlined by BS10, the semi-implicit integrator shares the same structure as the  Boris pusher and, as such, it preserves geometrical orbital properties.
This has the desirable properties to produce energy errors which remain bounded in time. 

\subsubsection{The Exponential Midpoint Method}
\label{sec:exp_midpoint}
The second method we consider belongs to the class of exponential integrators, a particular class of numerical schemes for integrating  stiff or highly oscillatory systems of ordinary differential equations (ODEs) see, for instance, \cite{Cox_Matthews_2002, HO_2010} and more recently \cite{Shen_Leok.2019}.
Our motivation for introducing exponential integrators dwells in the possibility of overcoming the limitations imposed by traditional explicit methods when tackling stiff sets of ODEs.
The stiffness is usually caused by a Jacobian that possesses eigenvalues with large negative real parts.
The prototype equation which is commonly adopted is
\begin{equation}\label{eq:etd_prototype}
  \frac{dv}{dt} = \tens{A}v + G(t,v)  \,,
\end{equation}
where $v$ and $G$ are vectors $\in\mathbb{R}^n$ while $\tens{A}$ is a square $n\times n$ constant-coefficient matrix usually with large spectral radius and thus potentially leading to stiffness.
Upon multiplying both sides by $e^{-\tens{A}t}$ and integrating between $t^n$ and $t^{n+1} = t^n + \Delta t$, one arrives at the variation of constant formula,
\begin{equation}\label{eq:etd_integral}
  v^{n+1} = e^{\tens{A}\Delta t} v^n
            + \int_0^{\Delta t} e^{\tens{A}(\Delta t-\tau)}
                   G\big(t^n+\tau, v(t^n+\tau)  \big)\,d\tau \,.
\end{equation}
This formula is exact, and different levels of approximation can be used to evaluate the integral on the right hand side.
At the simplest level, the integral can be approximated by assuming $G\approx G^n$ to be constant and carry out the integration in the exponential function only.
This recovers the exponential Euler method:
\begin{equation}\label{eq:etd1}
  v^{n+1} = e^{\Delta t\tens{A}}v^n + \Delta t \varphi_1(\tens{A}\Delta t) G^n \,,
\end{equation}
where $\varphi_1(\tens{A}\Delta t)$ is the matrix
\begin{equation}\label{eq:phi1}
  \varphi_1(\tens{A}\Delta t) = (\tens{A}\Delta t)^{-1}
                                (e^{\tens{A}\Delta t}-1) \,.
\end{equation}

Note that the exponential Euler method is first-order accurate when $G$ is a nonlinear function but it actually gives the \emph{exact} solution when $G$ is constant.
The exponential term $e^{\tens{A}\Delta t}v^n$ captures the stiff behavior exactly thus removing numerical instabilities possibly bound to the linear term.

Since the drag term is linear in the velocity, the particle equation of motion (\ref{eq:dust_ode}) has essentially the same form as the prototype ODE (\ref{eq:etd_prototype}) and a higher-order approximation can be obtained by evaluating $G$ at the half time level in Eq. (\ref{eq:etd_integral}).
This yields the \emph{exponential midpoint} rule which replaces the velocity update during the kick step.
Thus our exponential midpoint (EM) particle pusher can be outlined as follows:
\begin{equation}\label{eq:exp_midpoint}
  \begin{array}{lcl}
    \vec{x}^{n+\HALF} &=&\DS \vec{x}^n + \frac{\Delta t}{2}\vec{v}^{n} \,,
    \\ \noalign{\smallskip}
    \vec{v}^{n+1} &=&\DS   e^{\tens{A}\Delta t} \vec{v}^n
                  + h_1\vec{G}\left(\vec{x}^{n+\HALF},\,
                                    \vec{v}^{n+\HALF},\,
                                    \vec{v}_g^{n+\HALF} \right) \,,
    \\ \noalign{\smallskip}
    \vec{x}^{n+1} &=&\DS \vec{x}^{n+\HALF} + \frac{\Delta t}{2}\vec{v}^{n+1} \,,
  \end{array}
\end{equation}
\BLUE
where $\vec{v}^{n+\HALF} = (\vec{v}^n + \vec{v}^{n+1})/2$ while $h_1=\Delta t\varphi_1(\tens{A}\Delta t)$ is the exponential propagator.
Eq. (\ref{eq:exp_midpoint}) differs from Eq. (\ref{eq:semi-implicit}) only in the kick step which is realized using an exponential quadrature rule.
\BLACK
The exponential midpoint method has a local error of $\Delta t^3$ and it has globally second-order accuracy, just as the semi-implicit scheme.
Eq. (\ref{eq:exp_midpoint}) is time reversible and it has the correct asymptotic solution in the limit of constant $G$, where one should expect $\vec{v}_p \to -\tens{A}^{-1}\vec{G}$ (when $\tens{A}$ has real negative eigenvalues).
For infinitely large stopping times, $\tau_s\to\infty$, one recovers the semi-implicit method (\ref{eq:semi-implicit}) which preserves geometric orbital properties. 

\BLUE
In our implementation, we treat as stiff only the linear drag term so that, in Eq. (\ref{eq:etd_prototype}),  $\tens{A} = -{\rm diag}(1/\tau_s)$ is purely diagonal while the $\vec{G}$ vector contains the remaining terms\footnote{Note that, for the shearing-box equations, one could also include the Coriolis force during the linear step by introducing off-diagonal terms in the $\tens{A}$ matrix.
While this allows to retrieve the exact solution in the case of epicyclic motion, we did not find significant advantages in terms of stability and accuracy for the regimes investigated here.}.
Then the exponential propagator $h_1$ becomes, using Eq. (\ref{eq:phi1}):
\begin{equation}\label{eq:h1}
  h_1 = \Delta t\varphi_1\left(-\frac{\Delta t}{\tau_s}\right)
      = \tau_s(1-e^{-\Delta t/\tau_s}) \,.
\end{equation}

In the only presence of a drag force, for instance, $\vec{G}=\vec{v}_g/\tau$ and the kick step in (\ref{eq:exp_midpoint}) simplifies to 
\begin{equation}\label{eq:kick_modified}
  \vec{v}^{n+1} = e^{-\Delta t/\tau_s}\vec{v}^n
                    + h_1\frac{\vec{v}^{n+\HALF}_g}{\tau_s} \,.
\end{equation}

On the other hand, when $\vec{G}$ depends also on the particle velocity, the kick step becomes implicit and one has to invert the system for the velocity components.
However, for the equations' systems at hand, the inversion can be carried out analytically since $\vec{G}$ is either linear (in the case of the shearing-box equations) or can be obtained by updating the velocity components in a specific order.
\BLACK
For the shearing-box equations, the kick step is given in Appendix \ref{app:exp_midpoint_SB} while the extension to cylindrical and spherical coordinates is \BLUE given in the next section and in Appendix \ref{app:exp_midpoint_curv}.

The employment of exponential / exact integrators has also been introduced by \cite{Yang_Johansen.2016} who used operator (Strang) splitting for the mutual drag term and then resort to analytical integrations to relieve the time step constraint.
\BLUE
\subsection{Generalization to Curvilinear Coordinates}
\label{sec:curv_coordinates}
%

In the case of curvilinear geometries the derivation of the semi-implicit and exponential midpoint methods becomes slightly more involved as the right hand side of the particle ODE involves nonlinear expressions in the velocity and position (Eq. \ref{eq:dust_cylindrical} and \ref{eq:dust_spherical}).
Following the book of \cite{HLW_Book.2006}, we recall that for a generic Hamiltonian ${\cal H}(\vec{p},\vec{q})$, where $(\vec{q},\vec{p})$ are the canonical coordinates, the classical St\"ormer-Verlet method can be written as
\begin{equation}\label{eq:stormer_verlet}
  \begin{array}{l}
    \DS \vec{q}^{n+\HALF} = \vec{q}^n + \frac{\Delta t}{2}\nabla_{\vec{p}}
                            {\cal H}(\vec{p}^n, \vec{q}^{n+\HALF}) \,,
    \\ \noalign{\smallskip}
    \DS \vec{p}^{n+1}     = \vec{p}^n - \frac{\Delta t}{2}\Big(
                   \nabla_{\vec{q}} {\cal H}(\vec{p}^n, \vec{q}^{n+\HALF})
                 + \nabla_{\vec{q}} {\cal H}(\vec{p}^{n+1}, \vec{q}^{n+\HALF})
                  \Big) \,,
    \\ \noalign{\smallskip}
    \DS \vec{q}^{n+1} = \vec{q}^{n+\HALF} + \frac{\Delta t}{2}
                        \nabla_{\vec{p}} {\cal H}(\vec{p}^{n+1}, \vec{q}^{n+\HALF}) \,,
  \end{array}
\end{equation}
which is in the form drift-kick-drift.
In Eq. (\ref{eq:stormer_verlet}), $\vec{q} = (r,\,\phi,\,z)$ $\vec{p} = (v_r,\, l_\phi,\, v_z)$ are the canonical cylindrical coordinates in cylindrical geometry while $\vec{q} = (r,\, \theta,\, \phi)$, $\vec{p} = (v_r,\, l_\theta,\, l_\phi)$ in spherical coordinates.
Notice that the spatial coordinates $\vec{q}$ are always evaluated at the interval midpoint on the right hand side of Eq. (\ref{eq:stormer_verlet}).
The St\"ormer-Verlet method is a \quotes{geometrical integrator} in that it is symplectic and thus suitable for time-reversible, long-time integration.
For conservative mechanical system, the energy of the system oscillates around the expected (constant) value.
In absence of damping, the expression for the Hamiltonian may be found in classical mechanics textbooks.

Conversely, in presence of the viscous drag term, we modify the kick step of the exponential midpoint method consistently with Eq. (\ref{eq:exp_midpoint}), where $\vec{G}$ is now
\begin{equation}
  \vec{G}^{n+\HALF} = -\frac{ \nabla_{\vec{q}} {\cal H}(\vec{p}^n, \vec{q}^{n+\HALF})
                         + \nabla_{\vec{q}} {\cal H}(\vec{p}^{n+1}, \vec{q}^{n+\HALF})}
                              {2} + \frac{\vec{p}^{n+\HALF}_g}{\tau_s} \,.
\end{equation}
where $\vec{p}_g$ is the gas momentum vector a the half-time level.
The explicit expressions of Eq. (\ref{eq:stormer_verlet}) in cylindrical and spherical coordinates are reported, for lengthy reasons, in Appendix \ref{app:exp_midpoint_curv}.

\BLACK

\subsection{Connection between Grid and Particle Quantities}
\label{sec:weights}
%
The formalism employed in paper I will be used to establish the connection between particles and grid quantities.
In particular, deposition is used to collect and transfer particle attributes on the grid:
\begin{equation}\label{eq:deposition}
  Q_\indx = \sum_{p=1}^{N_p} W(\vec{x}_{\indx}-\vec{x}_p) q_p \,,
\end{equation}
\BLUE
where $q_p$ is typically the particle mass or force.
\BLACK
The function $W(\vec{x}_{\indx}-\vec{x}_p)$ are the kernel (or weight) functions.
Following paper I and the general framework used in PIC codes \citep[see Section 3.3 of Paper I and the book by][]{Birsdall_and_Langdon.2004}, we define $W$ as $W(x_i - x_p)W(y_j - y_p)W(z_k - z_p)$, i.e., the product of three one-dimensional weight functions.
The weights \BLUE specify which fraction of the particle shape overlaps with zone $\indx$: \BLACK
\begin{equation}\label{eq:weight}
  W_i \equiv W(\vec{x}_\indx - \vec{x}_p) = \int_\indx S(\vec{x} - \vec{x}_p)d^3x \,.
\end{equation}
The definition of the shape function is based on $b-$splines of increasingly higher order \citep[see, for instance,][]{Lapenta.2014} so that,
\begin{equation}\label{eq:shape_cart}
  S(\vec{x} - \vec{x}_p) = \frac{1}{\Delta x_1\Delta x_2\Delta x_3} \prod_{d=1}^{d=3}
                            b_m\left(\frac{x_d - x_{p,d}}{\Delta x_d}\right) \,,
\end{equation}
where $m$ is the spline order, $(x_1,x_2,x_3) = (x,y,z)$  and, for consistency, it is required that
\begin{equation}\label{eq:shape_norm}
  \int_{-\infty}^{+\infty} S(\vec{x} - \vec{x}_p) d^3x = 1  \,.
\end{equation}

In the present work we consider spline of order $0$ and $1$  respectively given by
\begin{equation}\label{eq:b0}
  b_0(\delta) = \left\{\begin{array}{ll}
      1   & \DS {\rm if} \quad |\delta| < \HALF  \\ \noalign{\smallskip}
      0   & \DS {\rm otherwise}
  \end{array}\right.      
\end{equation}
and
\begin{equation}\label{eq:b1}
  b_1(\delta) = \left\{\begin{array}{ll}
      1-|\delta|   & \DS {\rm if} \quad |\delta| < 1  \\ \noalign{\smallskip}
      0   & \DS {\rm otherwise}
  \end{array}\right.      
\end{equation}
which give a particle support that extend over no more than $3$ computational zones,
\BLUE
The weighting factors (Eq. \ref{eq:weight}) obtained by integrating Eq. (\ref{eq:shape_cart}) with $b_0(\delta)$ and $b_1(\delta)$ are the cloud-in-cell (CIC) and triangular shape cloud (TSC).
\BLACK
\subsubsection{Weighting Factors in Curvilinear Coordinates}
%

%

Traditional particle weighting schemes have been designed for a fixed mesh spacing and may results in systematic errors and loss of conservation when employed on a curvilinear grid, \cite[see, for instance,][]{Ruyten.1993, Larson_etal.1995, Verboncoeur.2001}.
The extension to non-Cartesian geometries is usually achieved by either interpolating in the volume coordinate or using correcting factors to restore charge or mass conservation.
In the traditional volume-weighing, for instance, the arguments of the spline functions are replaced by the volume coordinates so that standard Cartesian-like weighting can be used by replacing the linear coordinate with the volume coordinate.
On a cylindrical radial grid, for instance,  
\BLUE
\begin{equation}\label{eq:CIC_vol}
  \begin{array}{ll}
    W_i     &= \DS \frac{R_{i+1}^2 - R_p^2}{R^2_{i+1} - R^2_i} \,,
    \\ \noalign{\smallskip}
    W_{i+1} &= \DS 1 - W_i \,,
  \end{array}
\end{equation}
\BLACK
where $R_p\in[R_i, R_{i+1}]$ is the particle radial coordinate.

Nevertheless, in our experience, we have found volume (or area) weighting to lead to large errors.
For this reason \BLUE and in order to benefit from smoother higher-order weighting, \BLACK we propose to modify the shape function (\ref{eq:shape_cart}) into
\begin{equation}\label{eq:shape_curv}
  S(\vec{x} - \vec{x}_p) = \frac{a}{\Delta V_p} \prod_{d=1}^{d=3}
                            b_m\left(\frac{x_d - x_{p,d}}{\Delta x_d}\right)
\end{equation}
where $\Delta V_p$ is the volume support of the computational particle, $a$ is a correction factor specifically introduced for normalization purposes.
The shape function is now defined in terms of the linear coordinates - e.g. $(\Delta x_1,\, \Delta x_2,\, \Delta x_3) = (\Delta r,\, \Delta\theta,\, \Delta\phi)$ in spherical coordinates - rather than the volume ones.

In cylindrical coordinates, for instance, $(x_1,x_2,x_3)=(R,\phi,z)$ while $\Delta V_p = R_p\Delta R\Delta\phi\Delta z$.
When Eq. (\ref{eq:shape_curv}) is used together with Eq. (\ref{eq:weight}) one obtains, for $m=0$,
\begin{equation}\label{eq:CIC_cylindrical}
  \begin{array}{ll}
  W_i       &=\DS \frac{\delta + 2\nu_i}{2\delta + 2\nu_i}
                  \;  (1 - |\delta|)
  \\ \noalign{\smallskip}
  W_{i\pm1} &=\DS \frac{\delta + 2\nu_i \pm 1}{2\delta + 2\nu_i}
                  \;\frac{|\delta| \pm \delta}{2}
  \end{array}                                 
\end{equation}
where $\delta = (R_p-R_i)/\Delta R \in[-1/2,1/2]$, $\nu_i = R_i/\Delta R$ and $i$ is the index of the zone hosting the particle.
Equation (\ref{eq:CIC_cylindrical}) extends the traditional Cloud-in-Cell (CIC) weight function to the cylindrical radial direction.
For zero curvature ($\nu_i\to\infty$), one recovers the Cartesian weights defined by Eq. (43) in paper I.
Likewise, for the b-spline of order one, \BLUE one obtains the TSC weighting factors \BLACK
\begin{equation}\label{eq:TSC_cylindrical}
  \begin{array}{ll}
  W_i       &=\DS \frac{\delta + 3\nu_i}{3\delta + 3\nu_i}
                  \;\left(\frac{3}{4} - \delta^2\right)  \,,
  \\ \noalign{\smallskip}
  W_{i\pm1} &=\DS  \frac{\delta + 3\nu_i \pm 2}{3\delta + 3\nu_i}
                   \;\frac{1}{2}\left(\frac{1}{2}\pm\delta\right)^2 \,,
  \end{array}                                 
\end{equation}
which, in the limit of zero-curvature, \BLUE reduce to the corresponding expressions in Cartesian coordinates  (Eq. 44 of paper I).
\BLUE
It can be verified that $\sum_i W_i = 1$, as required by Eq. (\ref{eq:shape_norm}).
As we shall see in \S\ref{sec:rigid_disk}, the choice of a high-order shape function can considerably reduce the amount of grid noise which is inevitably introduced by unevenly spaced particles sampling a non-constant mass distribution.

We point out that the crucial difference between our CIC or TSC scheme and the traditional volume weighting is that Eq. (\ref{eq:CIC_cylindrical}) or (\ref{eq:TSC_cylindrical}) define the volume fraction of the particle shape occupying zone $\indx$ while Eq. (\ref{eq:CIC_vol}) is a simple interpolation in the volume coordinate.
In this sense, Eqs. (\ref{eq:CIC_cylindrical}) and  (\ref{eq:TSC_cylindrical}) may also be interpreted as \quotes{sliding} averages as the particles moves through the grid.
\BLACK
The corresponding expressions in spherical coordinates are more lengthy and are reported in Appendix (\ref{app:weights_spherical}).





\subsection{Time Step Determination}
%

PLUTO employs a dynamical time step estimated from the most recent time level,
\begin{equation}
  \Delta t^{n+1} = \min\Big(\Delta t^n_h, \Delta t^n_\D\Big)\,,
\end{equation}
where $\Delta t^n_h$ and $\Delta t^n_\D$ are the hydrodynamical and dust time steps, respectively.

The hydrodynamical step is computed from the Courant number, defined as 
\begin{equation}\label{eq:Courant}
  C_a = \left\{\begin{array}{ll}
   \DS \Delta t_h^n
        \max_{d,\indx}\left(\frac{|\lambda_{d,\indx}|}{\Delta l_{d,\indx}}\right)
   & \quad {\rm (CTU)} \,, \\ \noalign{\smallskip}
   \DS \Delta t_h^n\max_\indx
   \left(\frac{1}{N_{\rm dim}}\sum_d
   \frac{|\lambda_{d,\indx}|}{\Delta l_{d,\indx}}\right) 
   & \quad {\rm (RK2)}\,.
  \end{array}\right.
\end{equation}
In the expression above, $\lambda_{d,\indx}$ and $\Delta l_{d,\indx}$ are the maximum characteristic signal velocity and the cell length in zone $\indx\equiv(i,j,k)$ along the $d$ direction, respectively.
For stability, the Courant factor must respect the condition $C_a < 1$ (in 2D) or $C_a < 1/2$ (in 3D) for the CTU scheme while $C_a < 1/N_{\rm dim}$ for the RK2 scheme.
Here $N_{\rm dim}$ represents the number of spatial dimensions.
Eq. (\ref{eq:Courant}) does not consider dissipative terms which may be found on the PLUTO user's guide.

For the dust component, we prevent particles from traveling across more than 2 grid cells in order to avoid escaping of grains from the local processor domain in a single time step.
Similarly to paper I (see Eq. 33), this condition is expressed by:
\begin{equation}\label{eq:particles_dt}
  \Delta t^{-1}_{\D} = \max_{p,d}\left(
        \frac{|\hvec{e}_d\cdot\vec{v}_{p}^{n+\HALF}|}{N_{\rm max}\Delta l_d}
        \right) \,,
\end{equation}
where $N_{\rm max} = 1.8$ and the maximum extends to all particles and directions.
Note that no additional time step limitation is imposed from the stiffness of the equations when the exponential midpoint integrator is employed.

\section{Numerical Benchmark and Code Performance}
\label{sec:tests}
%
%
%
%

In this section we demonstrate the accuracy and robustness of our gas-particle hybrid scheme through a series of numerical benchmarks in Cartesian and curvilinear geometries.

\subsection{Particle-Gas Deceleration Test}
%

\begin{figure}[!ht]
  \centering
  \includegraphics[width=0.5\textwidth]{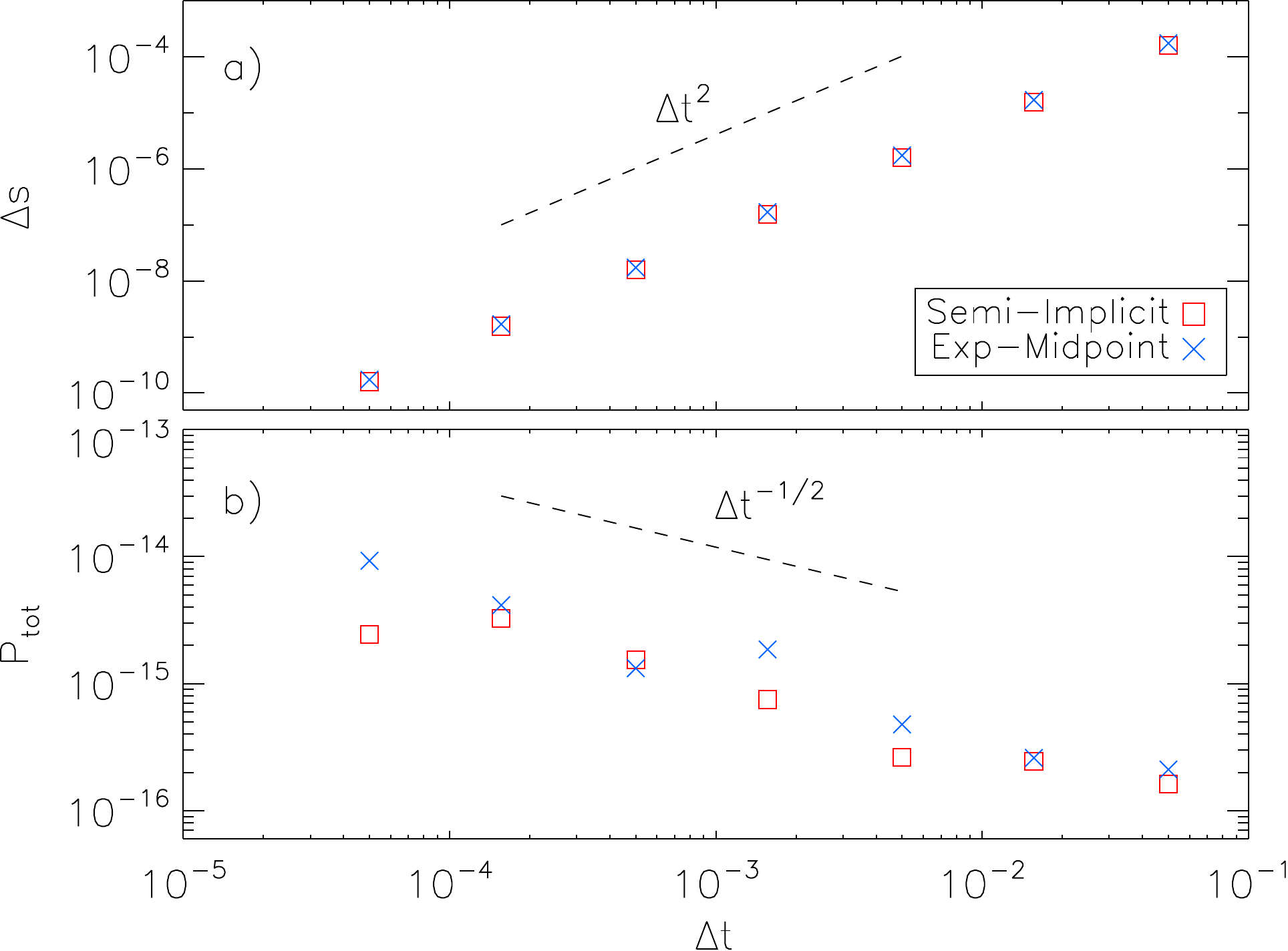}
  \caption{\footnotesize
           Temporal resolution study for the deceleration test problem.
           Errors in the particle position are plotted in the top panel as
           functions of the resolution for the semi-implicit (red squares)
           and the exponential midpoint (blue x) methods.
           In the bottom panel we show the total (gas+particle) momentum.
           \label{fig:deceleration1}}
\end{figure}

Following BS10, we investigate the dynamical interaction of an ensemble of uniformly distributed dust particles feeding back onto a homogeneous flow.
Since the system does not has spatial variations, gas and particle equations reduces to the simple system of ODE
\begin{equation}\label{eq:deceleration}
    \left\{\begin{array}{lcl}
    \DS \frac{d\vec{v}_g}{dt} &=&\DS  \epsilon
                                 \frac{\vec{v}_p - \vec{v}_g}{\tau_s} \,,
    \\ \noalign{\smallskip}
    \DS \frac{d\vec{v}_p}{dt} &=&\DS -\frac{\vec{v}_p - \vec{v}_g}{\tau_s} \,,
  \end{array}\right.
\end{equation}
where $\epsilon$ is the particle-to-gas mass ratio.
In the center of mass frame, where $\epsilon \vec{v}_p(0) + \vec{v}_g=\vec{0}$, the solution of (\ref{eq:deceleration}) can be obtained exactly as:
\begin{equation}\label{eq:deceleration_solution}
    \left\{\begin{array}{lcl}
    \vec{v}_p(t) &=&\DS  \vec{v}_p(0)e^{-(1+\epsilon)t/\tau_s} \,,
    \\ \noalign{\smallskip}
    \vec{x}_p(t) &=&\DS \vec{x}_p(0)
                    + \frac{\tau_s\vec{v}_p(0)}{1+\epsilon}
                        \left[1-e^{-(1+\epsilon)t/\tau_s}\right] \,.
                       
    \end{array}\right.
\end{equation}
\BLUE
For the present test we set $\rho=1$, $\epsilon = 1$ \BLACK and consider motion along the $x$ direction, $\vec{v}_p = -\vec{v}_g = (1,0,0)$ using one particle per cell.

We first conduct a temporal resolution study using $\tau_s = 2$ and logarithmically increasing time step sizes, $\Delta t = \Delta t_0 10^{n/2}$ where $\Delta t_0 = 5\times 10^{-5}$ and $n=0,1...6$.
Cumulative position errors are computed as ${\rm err} = (\min_p({\rm err}_p)+\max_p({\rm err}_p))/2$ where
\begin{equation}
  {\rm err}_p = |x_p(t_f) - x_p^{\rm ex}(t_f)|
\end{equation}
is the deviation of the $p$-th particle position with respect to the analytic solution (Eq. \ref{eq:deceleration_solution}) at the final integration time $t_f=1$.
Errors obtained using the semi-implicit and exponential schemes are compared in the top panel of Fig.~\ref{fig:deceleration1}: both methods achieve second-order accuracy (as expected) with evenly matched errors.
The x-component of the total (particle+gas) momentum is plotted, still as a function of the time step, in the bottom panel of Fig.~\ref{fig:deceleration1}.
Although the total momentum should remain zero, fluctuations at the machine-accuracy level are observed with $\sim 1/\sqrt{\Delta t}$ scaling indicating randomly and uncorrelated events.

\begin{figure}
  \centering
  \includegraphics[width=0.5\textwidth]{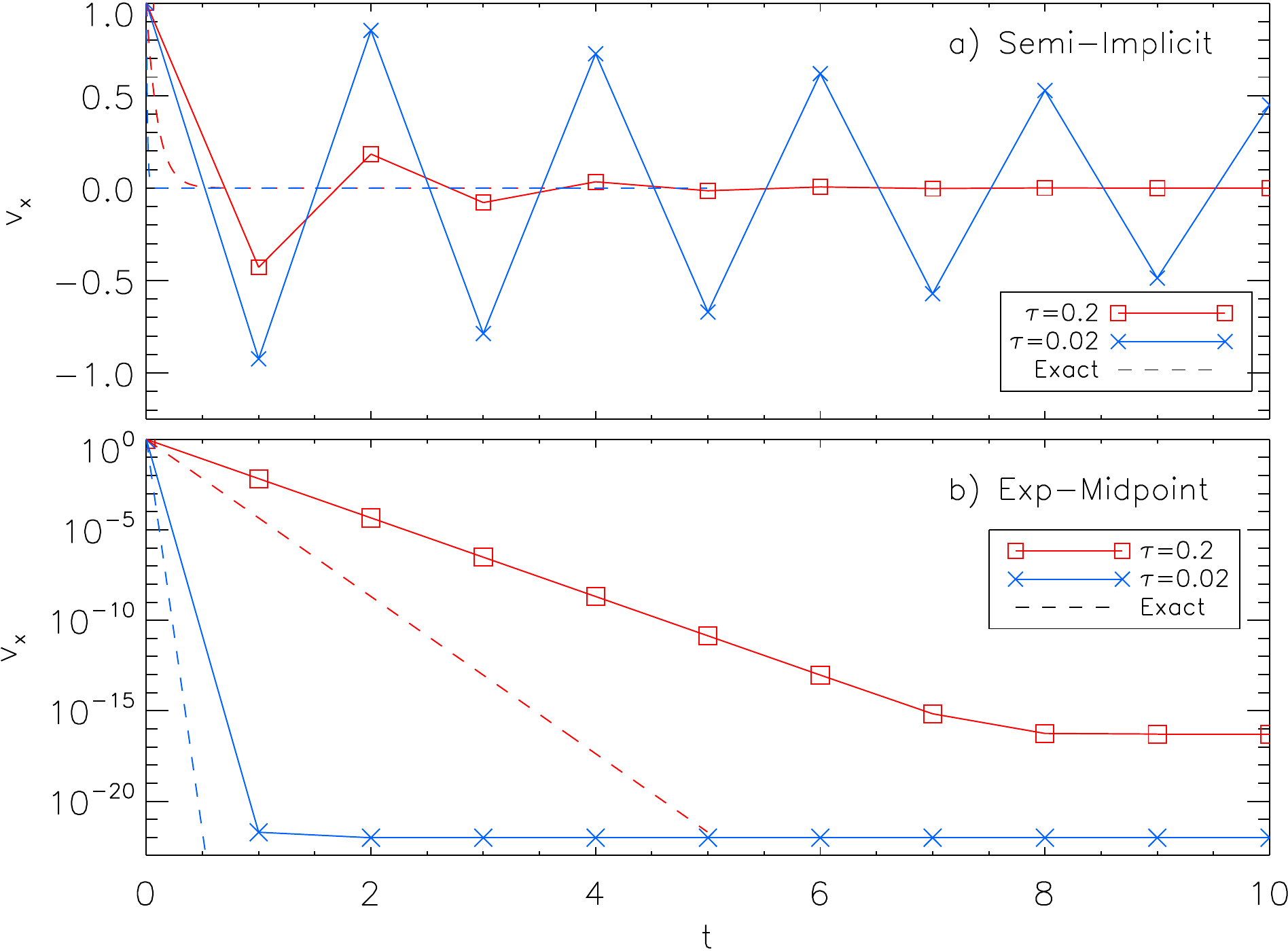}
  \caption{\footnotesize
           Comparison between the semi-implicit method (top panel) and the
           exponential midpoint method (bottom panel) for the deceleration test.
           In each panel we plot the particle velocity as a function of time
           with red and blue lines corresponding to $\tau_s=0.2$ (squares)
           and $\tau_s=0.02$ (x symbols), respectively.
           Dashed lines of the same color give the exact solution.
           Note that the solution obtained with the exponential midpoint rule
           (bottom) is completely oscillation-free and plotted using log scale.
           \label{fig:deceleration2}}
\end{figure}
Next we assess the algorithms performance in the limit of stiff drag force.
We employ $\tau_s = 0.2$ and $\tau_s = 0.02$ and integrate the fluid-dust equations using a constant time increment $\Delta t = 1$ until $t_f=10$.  
The particle velocity is plotted as a function of time in Fig.~\ref{fig:deceleration2} for the semi-implicit (top) and exponential midpoint (bottom) methods.
Particles with stopping times $\tau_s=0.2$ and $\tau_s=0.02$ are shown using red squares
blue x points, respectively.
Note that the time step has been chosen to exceed the particle stopping time and hence the damping rate cannot be captured accurately.
While our results agree well with those of BS10 in the case of the semi-implicit scheme (see their figure 3), our exponential integrator clearly outperforms both the semi-implicit and the fully implicit scheme of BS10 yielding a more accurate evolution.
More precisely, the solution decays rapidly in $\sim 8$ steps when $\tau_s = 0.2$ and even more rapidly for $\tau_s=0.02$ with velocity.
Even if the actual stopping time is completely under-resolved, the particle velocity \BLUE reaches the value of $\approx 10^{-22}$ \BLACK in a single step thus approaching the theoretical value more closely.
On the other hand, the semi-implicit method shows an oscillatory behavior with reduced damping (in contrast to the prediction) as $\tau_s$ is decreased.
This is potentially dangerous since resonant interactions with the fluid may potentially amplify the oscillation amplitudes.
The $L_1$-norm errors for the semi-implicit scheme are $\epsilon_1(v_x)\approx 0.068$ (for $\tau_s=0.2$) and $\epsilon_1(v_x) \approx 0.6$ (when $\tau_s=0.02$) while, in the case of the exponential rule, we obtain $\epsilon_1(v_x) = 6.1\times10^{-4}$ and $\epsilon_1(v_x) = 9.6\times 10^{-23}$. 
We thus conclude that while the exponential method yields excellent results in both the non-stiff and stiff regimes, the semi-implicit scheme can be safely employed only when $\Delta t \gtrsim \tau_s$.

\begin{figure*}[!ht]
  \centering
  \includegraphics[width=0.75\textwidth]{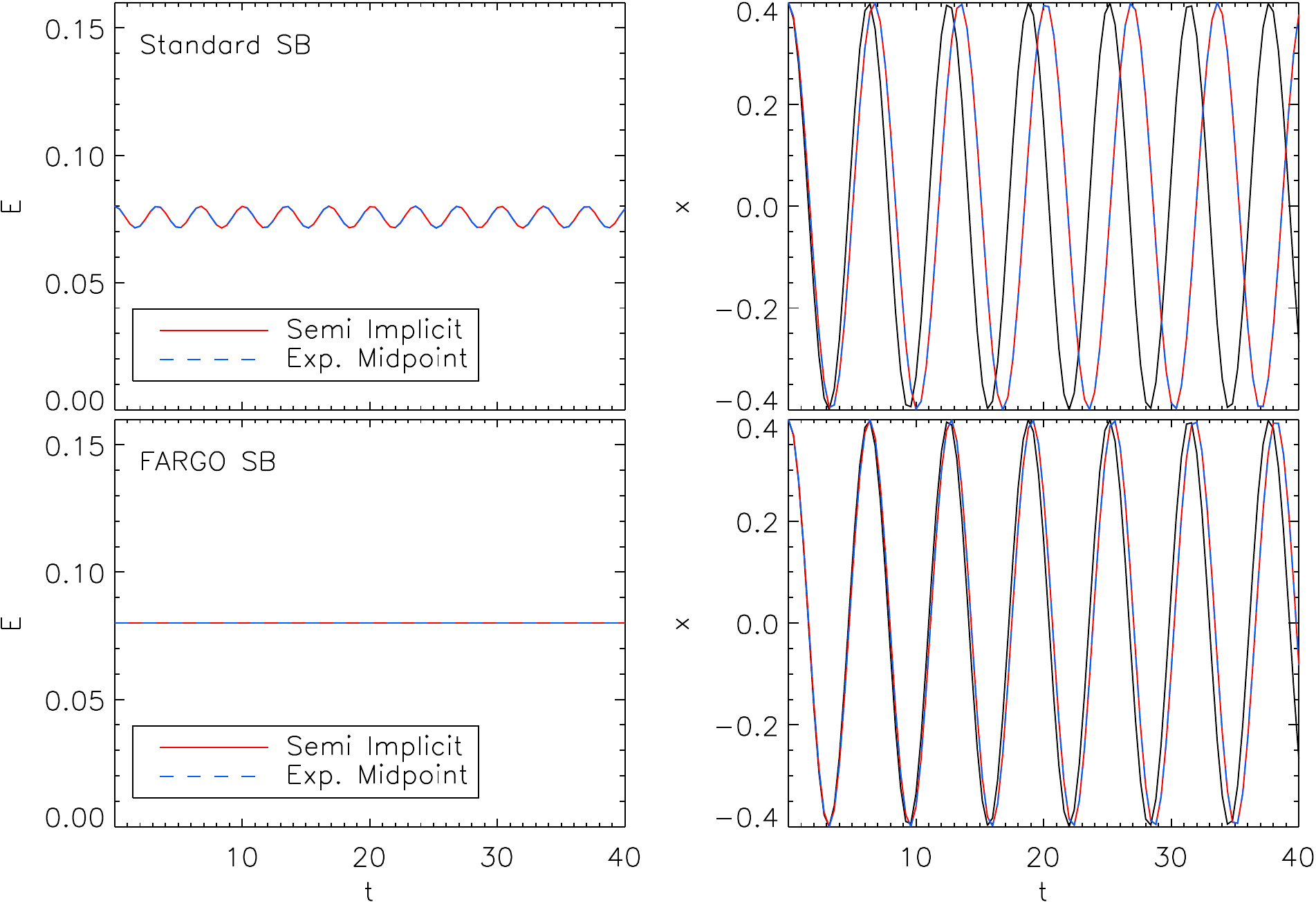}
  \caption{\footnotesize
           Results for the epicyclic motion test problem using the standard
           shearingbox equations (top panels) and with the orbital advection
           scheme (FARGO, bottom panels).
           The left and right panels show, respectively, the time evolution
           of the energy and of the $x$-coordinate.
           Red and blue lines denotes the numerical solution obtained with
           the semi-implicit and the exponential methods while the black
           line in the right panels corresponds to the exact analytic solution
           given by Eq. (\ref{eq:epicyclic}).
           \label{fig:epicyclic}}
\end{figure*}

\subsection{Epicyclic Motion}
%

In the next problem we solve for the motion of a single dust particle in the local shearing sheet approximation, ignoring the drag force ($\tau_s = \infty$).
We elaborate on the same test presented in BS10 and solve the particle shearingbox equations using the standard formulation (Eq. \ref{eq:standard_SB_g} and \ref{eq:standard_SB_p}) as well as the orbital advection scheme (Eq. \ref{eq:FARGO_SB_g} and \ref{eq:FARGO_SB_p}).
In the standard SB frame, the particle initial condition is specified by $\vec{x}_p(0) = (A,0,0)$ and $\vec{v}_p(0) = (0,-2\Omega A,\, 0)$ where $A = 0.4$, $\Omega = 1$.
The solutions can be obtained analytically giving 
\begin{equation}\label{eq:epicyclic}
  \left\{\begin{array}{lcl}
    \vec{x}_p(t) &=&\DS  A\left[ \cos (\omega t),\,
                                -\frac{2\Omega}{\omega}\sin(\omega t),\,
                                0\right]
  \\ \noalign{\smallskip}
  \vec{v}_p(t) &=& \DS -\omega A\left[\sin(\omega t),\,
                                \frac{2\Omega}{\omega}\cos (\omega t),\,
                                0 \right] 
  \end{array}\right.
\end{equation}
where $\omega = \Omega\sqrt{4-2q}$ is epicyclic frequency.
The particle orbit is thus an ellipse with axes $(1/A)$ and $\omega/(2\Omega A)$.
The epicyclic motion conserves total energy, viz.
\begin{equation}\label{eq:epicyclic_energy}
  E_p = \frac{1}{2}\vec{v}_p^2 - q\Omega^2 x_p^2 = \frac{\omega^2 A^2}{2} \,,
\end{equation}
which will be useful to assess the conservation properties of our scheme.

Computations are carried out up to $t_f = 40$ using a fixed time step of $\Delta t = 0.4/\Omega$ deliberately chosen to be large in order to emphasize the numerical error.
Results are plotted in the top panels of Fig.~\ref{fig:epicyclic} (using the standard shearing-sheet equations) and in the bottom panels of the same figure (using the orbital advection scheme).
Left and right panels show, respectively, the particle energy and the $x$-coordinate as a function of time.
Computations have been compared using the semi-implicit (red curves) as well as the exponential-midpoint (blue curves) methods.

The two different time-stepping schemes produce identical results in this limit ($1/\tau_s \approx 0$).
Within the standard shearingbox, energy oscillates around the exact value as expected for a time-reversible Leap-Frog-like integrator.
Owing to the large time step, the phase error is considerably pronounced.
Conversely, when using the orbital scheme, total energy is conserved exactly and phase error are noticeably reduced.
Notice that the typical time step employed in actual computations is much smaller than the one chosen here and so is the phase error.
\BLUE
Our results show that the exponential integrator perform as well as the semi-implicit integrator of BS10.
\BLACK

\subsection{Streaming Instability in the local Shearingbox Model}
%

\begin{figure*}
  \centering
  \includegraphics[width=0.8\textwidth]{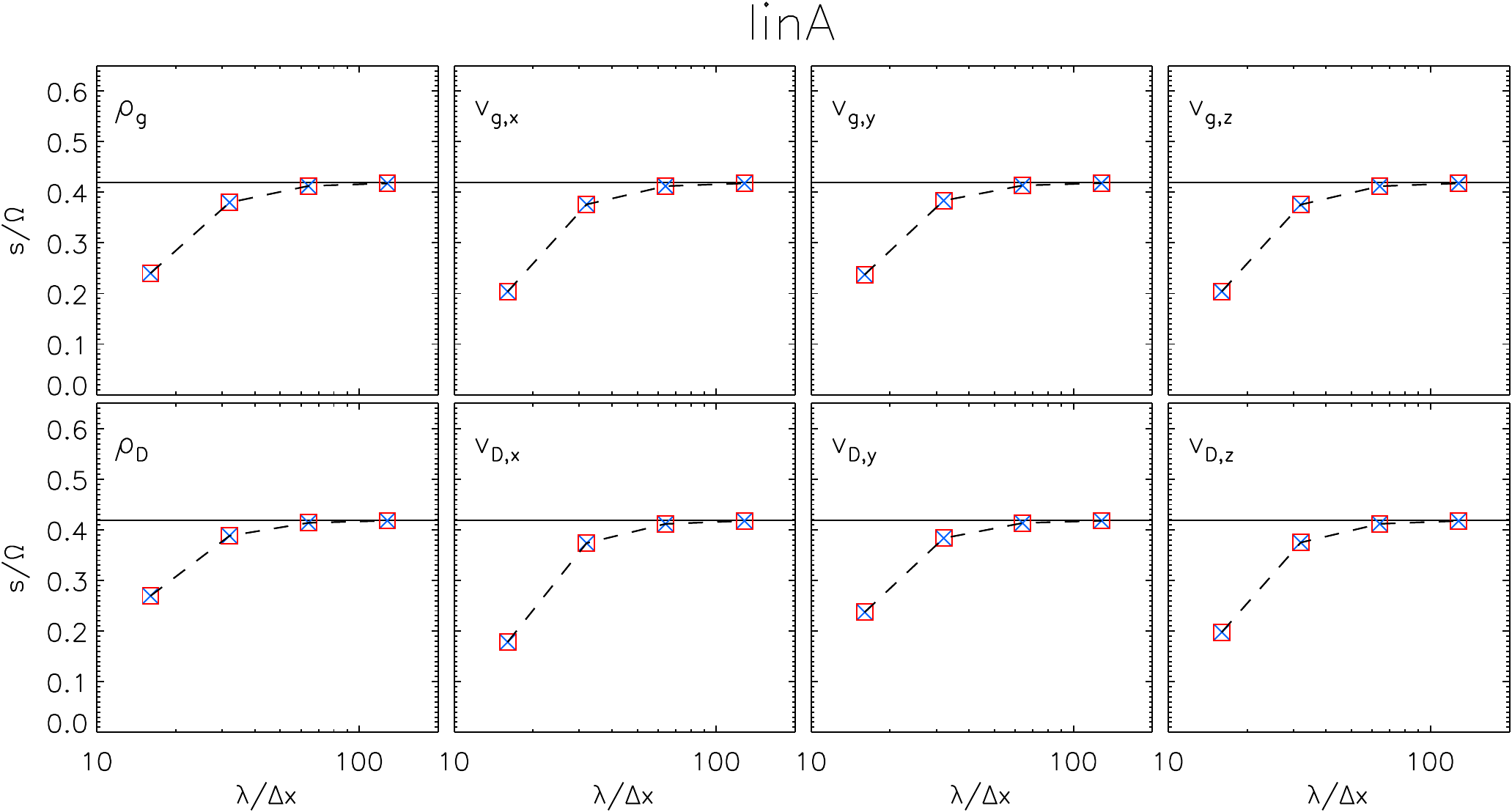}
  \includegraphics[width=0.8\textwidth]{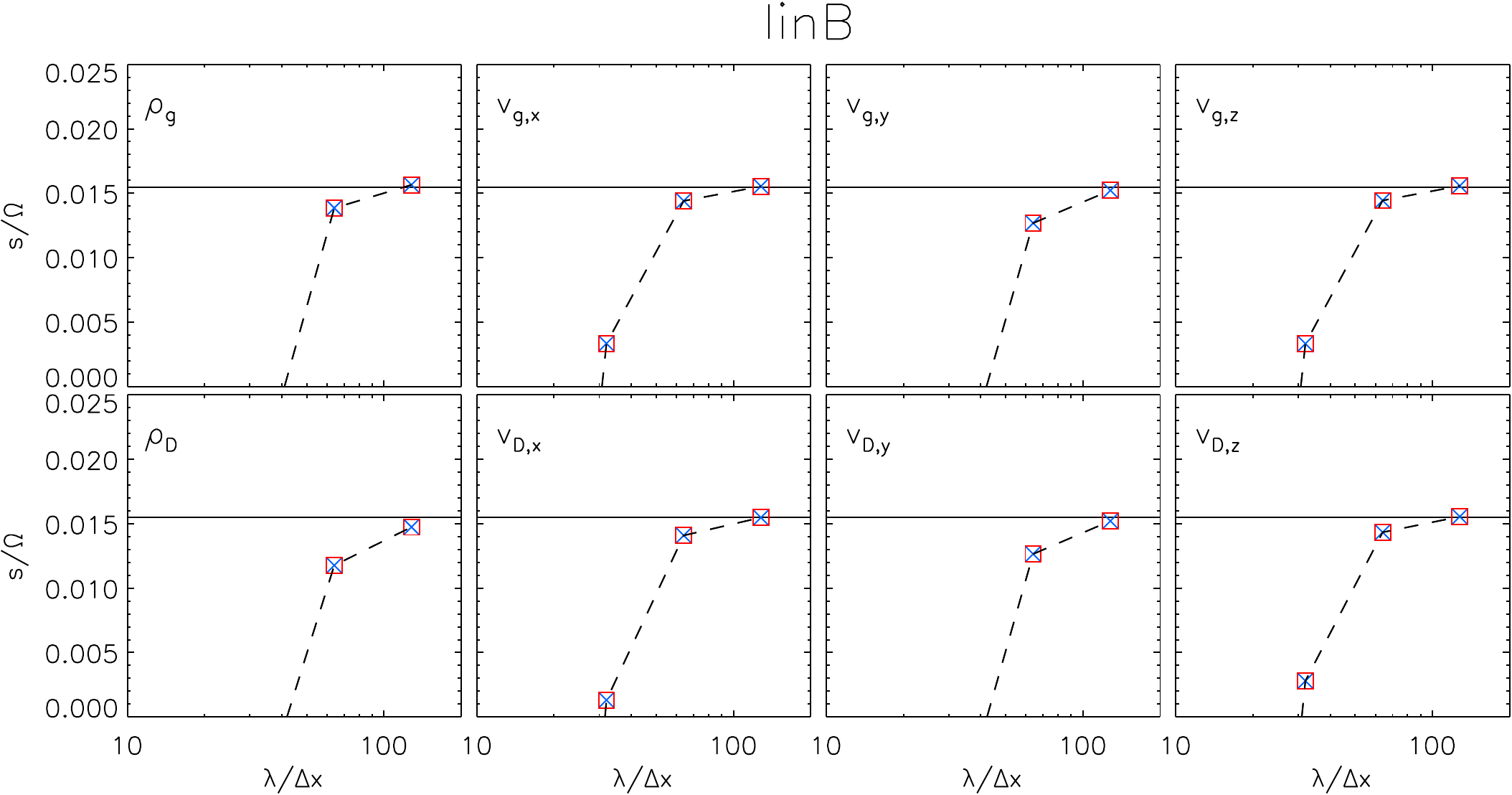}
  \caption{\footnotesize
           Results for the 2D streaming instability showing the computed
           numerical growth rate (red squares: semi-implicit method,
           blue crosses: exponential midpoint method) as a function of the
           number of grid points per wavelength.
           Since one wavelength is used, this number corresponds to the actual
           numerical resolution.
           Top and bottom plots corresponds to test \emph{linA} and \emph{linB},
           respectively.
           In each plot we show, from left to right, the growth rate measured
           on density and the three components of velocity.
           Analytical growth rates are shown as a solid black line.
           \label{fig:streaming_instability}}
\end{figure*}
Next we illustrate an application of the dust particle module to the dynamics of protoplanetary disks in the shearingbox approximation.
The initial condition consists of an equilibrium state between gas and solid components in an unstratified Keplerian disks as described, among others, in \cite{NSH.1986}, JY07, BS10.
Deviations from the Keplerian rotation are expressed, for the gas and particles, respectively by
\begin{equation} \label{eq:SI_equil}
  \begin{array}{ll}
  \vec{v}_g &= \DS \left[\frac{2\epsilon\tilde{\tau}_s}{\Delta}\hvec{e}_x
   -\left(1 + \frac{\epsilon \tilde{\tau}_s^2}{\Delta}\right)
                     \frac{1}{1+\epsilon}\hvec{e}_y\right]\eta v_K
  \\ \noalign{\smallskip}                      
  \vec{v}_p  &=\DS -\left[\frac{2\epsilon\tilde{\tau}_s}{\Delta}\hvec{e}_x
   +\left(1 - \frac{\tilde{\tau}_s^2}{\Delta}\right)
                     \frac{1}{1+\epsilon}\hvec{e}_y\right]\eta v_K
  \end{array}                     
\end{equation}
where $\epsilon$ is the density ratio of particles to gas, $\Delta = (1+\epsilon)^2 + \tau_s^2$, $\tilde{\tau}_s=\Omega\tau_s$ is the Stokes number and $\eta$ has been defined in Eq. (\ref{eq:eta}).
We employ $\Omega = 1$, $q = 3/2$.
%

For verification purposes, we set $\eta v_K = c_s/20$ and check our results against two benchmark configurations extensively employed by previous investigators \citep[see, e.g., YJ07, ][BS10]{Balsara_etal.2009}.
The linear stability problem is then uniquely defined in terms of four parameters \citep{YG.2005}, namely, the background dust to gas mass ratio, the particle Stokes number, and the dimensionless wave numbers in the radial and vertical directions, $K_x = k_x \eta v_K/\Omega$ and $K_z = k_z \eta v_K/\Omega$.
Choosing the box sizes $L_x = L_z$ to contain exactly one wavelength, the sound speed can be obtained from the dimensionless wavenumbers:
\begin{equation}
  c_s = \frac{20K_x\Omega L_x}{2\pi} \,.
\end{equation}

The system is perturbed using the exact eigenmodes of the linearized system which include modes of even ($e$) and odd ($o$) parities:
\begin{equation}
  \begin{array}{lcl}
  \delta Q_e &=&  A\left[  \Re(\tilde{Q})\cos(\phi_x)
                         - \Im(\tilde{Q})\sin(\phi_x)\right] \cos(k_zz)e^{st}
  \\ \noalign{\smallskip}
  \delta Q_o &=& -A\left[  \Re(\tilde{Q})\sin(\phi_x)
                         + \Im(\tilde{Q})\cos(\phi_x)\right] \sin(k_zz)e^{st}
  \end{array}
\end{equation}
where $Q$ represents a fluid or particle quantity, $\Re(\tilde{Q})$ and $\Im(\tilde{Q})$ are the real and imaginary components of the eigenvector, $\phi_x = k_xx-\omega_\Re t$ and $A$ is the perturbation amplitude which we initially set to $10^{-6}$.
The eigenvalue of the system are given by $\omega = \omega_\Re + {\rm i}s$.
Odd perturbation modes are used for the vertical velocities while even perturbations  are employed for the remaining variables, i.e., $Q = \{\rho, v_x, v_y\}$.
The explicit numeric values of the eigenmodes for the two configurations (\emph{linA} and \emph{linB}, respectively) are reported in Table 1 of YJ07.
In the first set of computations (case \emph{linA}), the dust to gas mass ratio, stopping time and dimensionless wave numbers are, respectively, set to $\epsilon = 3$, $\tau_s=0.1$ and $K_x = K_z = 30$.
The expected growth rate, for this configuration is $s = 0.4190204\, \Omega$.
The \emph{linB} case is defined by $\epsilon = 0.2$, $\tau_s=0.1$ and $K_x = K_z = 6$ and it is more severe owing to a much smaller growth rate, $s = 0.015476\, \Omega$.

In order to seed linear particle density perturbations, we adopt the same procedure illustrated in the appendix of YJ07.
Assuming eigenvector to be normalized to the particle density perturbation, the density profile of the dust component should behave as $\rho_{\D} = 1 + A\cos(k_xx)\cos(k_zz)$.
To approximate such a distribution we displace particles by an amount
\begin{equation}
  \xi_x = \frac{A}{k_x}\left[-\sin(k_xx) + \frac{A}{2}\sin(2k_xx)\right]
\end{equation}
relative to their cell center.
The second-order term in the displacement approximates the desired distribution up to $O(A^3)$ when used in conjunction with the TCS weighting scheme (see the appendix of YJ07).
This approach concentrates the power in the desired mode and reduce the stochastic noise on the grid.

Computations are performed on the square box $x\in[-L_x/2,L_x/2]$ and $z\in[-L_z/2,L_z/2]$ with $L_x=L_z=2$ by solving the axisymmetric shearingbox equations.
We employ the CTU-PPM scheme with fixed Courant number $C_a=0.8$.
Note that, with the orbital advection scheme, the boundary conditions in the radial ($x$) direction are simply periodic and the numerical integration can be carried in a standard Cartesian frame with non-inertial acceleration term given by Eq. (\ref{eq:FARGO_SB_g}).
We employ one particle per cell with equilibrium position corresponding to the zone center.

Growth rates are computed by first measuring the deviations from the equilibrium state,
\begin{equation}
  \delta Q^n = \max_\indx\left(Q^n - Q_{\rm eq}\right)  \quad{\rm for}\quad
  t\in[0,6] \,,
\end{equation}
where the maximum is taken over the entire computational domain and then by fitting the paired values of $\{t^n,\delta Q^n\}$ with a linear model by minimizing the chi-squared statistic.
Fig.~\ref{fig:streaming_instability} shows the computed growth rates for different spatial resolutions, $N_x = N_z = 16, 32, 64, 128$.
Top and bottom panels correspond to case \emph{linA} and \emph{linB}, respectively, and we use red squares (blue crosses) to indicate the computations obtained with the semi-implicit (exponential midpoint) schemes.
In particular, $\gtrsim 32$ zones per wavelength are necessary to capture the correct growth rate in test \emph{linA} while approximately twice the resolution (at least) is needed for test \emph{linB}.
These conclusions hold for the semi-implicit method as well as the exponential midpoint method for which calculations are nearly identical.
Given the $2^{\rm nd}$ order accuracy of the scheme, our results well agree with those of BS10 obtained with the Athena code.
On the contrary, faster convergence is reached in YJ07 who employed the Pencil code with $6^{\rm th}$-order accuracy.
We have also checked our results using 4 and 9 particles per cell and trivial differences have been observed during the linear stages.

\subsection{Orbital Test in Cylindrical Coordinates}
%

\begin{figure*}[!ht]
  \centering
  \includegraphics[width=0.85\textwidth]{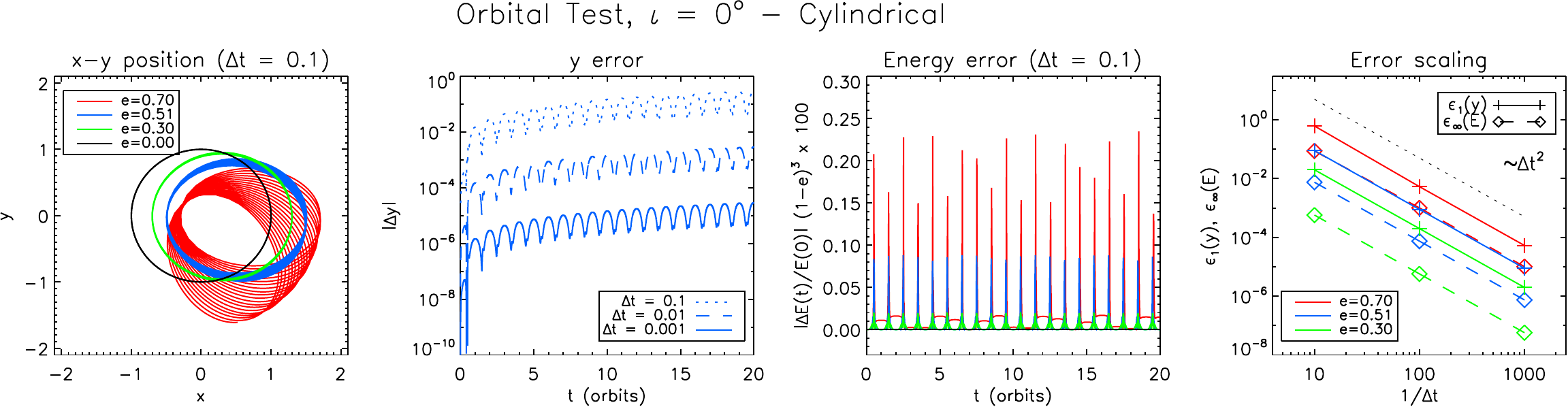}
  \caption{\footnotesize
\BLUE
           Orbital test in cylindrical coordinates using 4 particles with 
           eccentricities $e=0,\, 0.3,\, 0.51,\, 0.7$ (black, green, blue and red)
           and same orbital period $T = 2\pi a\sqrt{a}$ where $a=1$.
           Computations have been carried out for $20$ orbits and the inclination
           angle is $\iota = 0$.
           From left to right: i) trajectories in the $x-y$ plane for
           $\Delta t = 0.1$;
           ii) errors in the $y$ coordinates as a function of time
           for the particle with $e=0.51$ using different step sizes
           $\Delta t = 0.1, 0.01, 0.001$ (dashed, dotted and solid blue lines,
           respectively);
           iii) percent relative errors in energy as a function of time
           using $\Delta t = 0.1$ (a factor $(1-e)^3$ has been incorporated
           for scaling reasons);
           iv) $L_1$ norm errors for the $y$ coordinate (plus signs, solid lines)
           and $L_\infty$ for total energy (diamonds, dashed lines).
\BLACK
           \label{fig:orbital_test_i0}}
\end{figure*}

In order to assess the accuracy of our integrator in curvilinear coordinates we consider dust particles orbiting around a central object around elliptic trajectories.
We neglect the drag term and employ 4 different particles with eccentricities $e=(0, 0.3, 0.51, 0.7)$ and same semi-major axis $a=1$ so that the orbital period $T = 2\pi a^{3/2}$ is identical for all particles.

\BLUE We first consider particles initially placed on the midplane (zero inclination angle, $\iota=0$) at $\phi=0$ at a distance $R(0) = a(1-e)$ from the central object with angular momentum $v_\phi R = \sqrt{a(1-e^2)}$.
Particles are evolved for 20 orbits using the cylindrical integrator (Appendix \ref{app:EM_cylindrical}) with time step sizes $\Delta t = 0.1, 0.02, 0.04$.
The left panels in Fig.~\ref{fig:orbital_test_i0} show the trajectories computed with the largest time step ($\Delta t = 0.1$), in order to emphasize the error.
While integration is exact for a circular orbit, orbital precession becomes more pronounced for larger eccentricities.

As a measure of the error, we compute the difference $\Delta y(t) = y(t) - y_{\rm ref}(t)$ where $y(t)=R(t)\sin\phi(t)$ is the numerical solution and $y_{\rm ref}(t)=a\sqrt{1-e^2}\sin\psi(t)$ is obtained by solving, at each time step $t$, Kepler's equation
\begin{equation}
  f(\psi) = a\sqrt{a}(\psi - e\sin\psi) - t = 0
\end{equation}
with a Newton-Raphson method.
The absolute value of $\Delta y(t)$ is plotted as a function of time in the second panel of Fig.~\ref{fig:orbital_test_i0} for $e=0.51$ and for the selected time steps (dotted, dashed and solid blue lines).
Our results can be directly compared to and favorably agree with those of \cite[][see their figures 22 and 23]{Zhu_etal.2014}.

Similarly, we examine conservation of mechanical energy by plotting its relative error as a function of time in the third panel.
To emphasize the effect, we again show only the results with $\Delta t = 0.1$.
Our results reveal that errors increase with eccentricity but oscillate 
with the same periodicity of the system without increasing in time 
(indeed we have checked this property for hundreds of orbits).
Error scaling with time resolution is illustrated in the fourth panel, where the $L_1$ norm error of $|\Delta y|$ (solid, plus signs) and the $L_\infty$ error of energy (dashed, diamonds) are plotted confirming the second-order accuracy of the scheme.

\begin{figure*}[!h]
  \centering
  \includegraphics[width=0.85\textwidth]{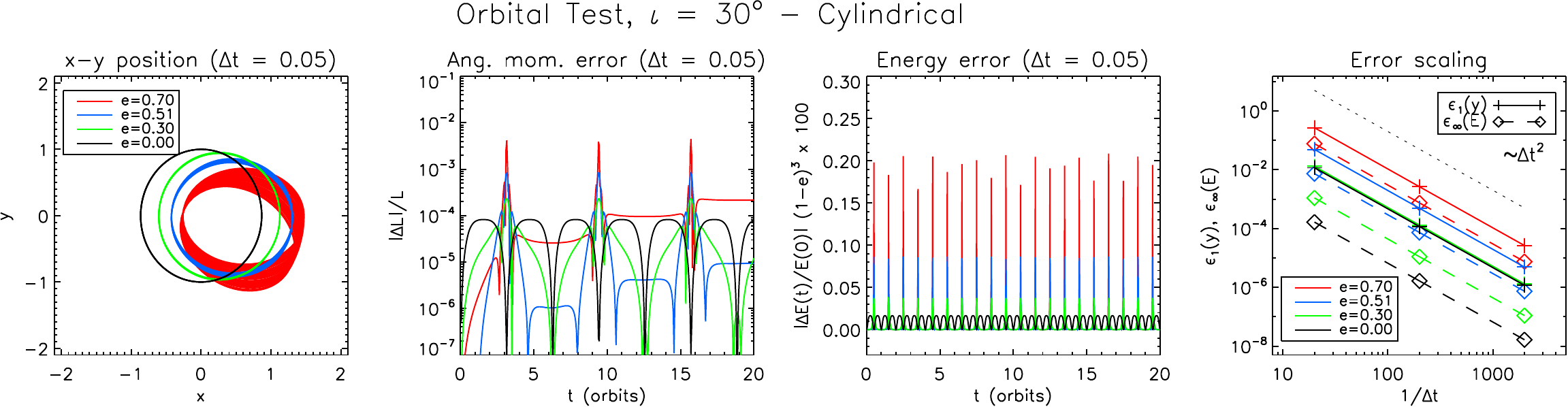}
  \includegraphics[width=0.85\textwidth]{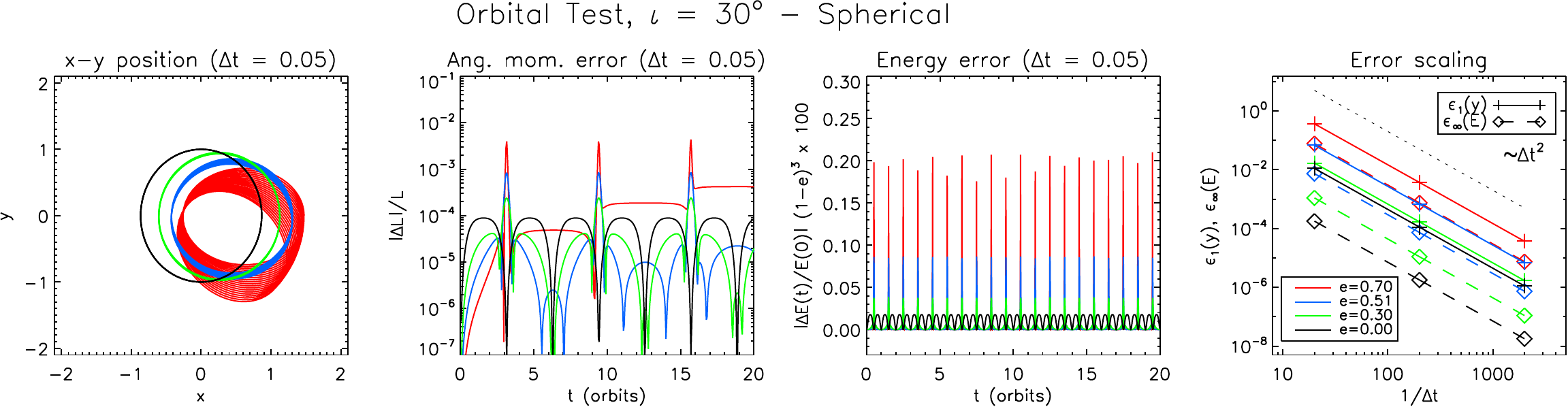}
  \caption{\footnotesize
\BLUE
           Orbital test for an inclination angle $\iota = 30^\circ$. 
           Top and bottom panels corresponds, respectively, to cylindrical and
           spherical computations.
           From left to right we show: i) projected orbits in the $x-y$ plane
           for different eccentricities; ii) relative error in total angular
           momentum $|\vec{L}|$;
           iii) percent relative errors in energy as a function of time
           using $\Delta t = 0.1$ (a factor $(1-e)^3$ has been incorporated
           for scaling reasons);
           iv) $L_1$ norm errors for the $y$ coordinate (plus signs, solid lines)
           and $L_\infty$ for total energy (diamonds, dashed lines).
           Results in the the first three panels are shown for $\Delta t = 0.05$
           to emphasize the error.
\BLACK
           \label{fig:orbital_test_i30}}
\end{figure*}
Next we repeat computations by considering tilted orbits, obtained by rotating the initial condition by an angle $\iota = 30^{\circ}$ around the $y$-axis.
Results in cylindrical and spherical coordinates are shown in the top and bottom panels of Fig. \ref{fig:orbital_test_i30}.
For stability, the time steps have been halved, that is, $\Delta t = 5\times(10^{-2},\, 10^{-3},\,10^{-3})$.
The leftmost panels now show the projected orbits on the midplane and, as in the previous case, spurious precession is enhanced at larger eccentricities.
Note that integration is no longer exact even in the case of a circular orbit and that the angular momentum vector $\vec{L} = (L_x, L_y, L_z)$ has non-vanishing components in all three directions.
Conservation of $\vec{L}$ is, however, within the truncation level of 
the scheme, as shown in the second pair of panels where the relative error of total angular momentum is plotted as a function of time.
Mechanical energy oscillates (as expected) around the nominal value for all eccentricities (third pair of panels) and error scaling computed as before confirm second-order convergence (fourth pair of panels).
\BLACK

\subsection{Radial Drift at different Stokes' Numbers}
%

\begin{figure*}
  \centering
  \includegraphics[width=0.75\textwidth]{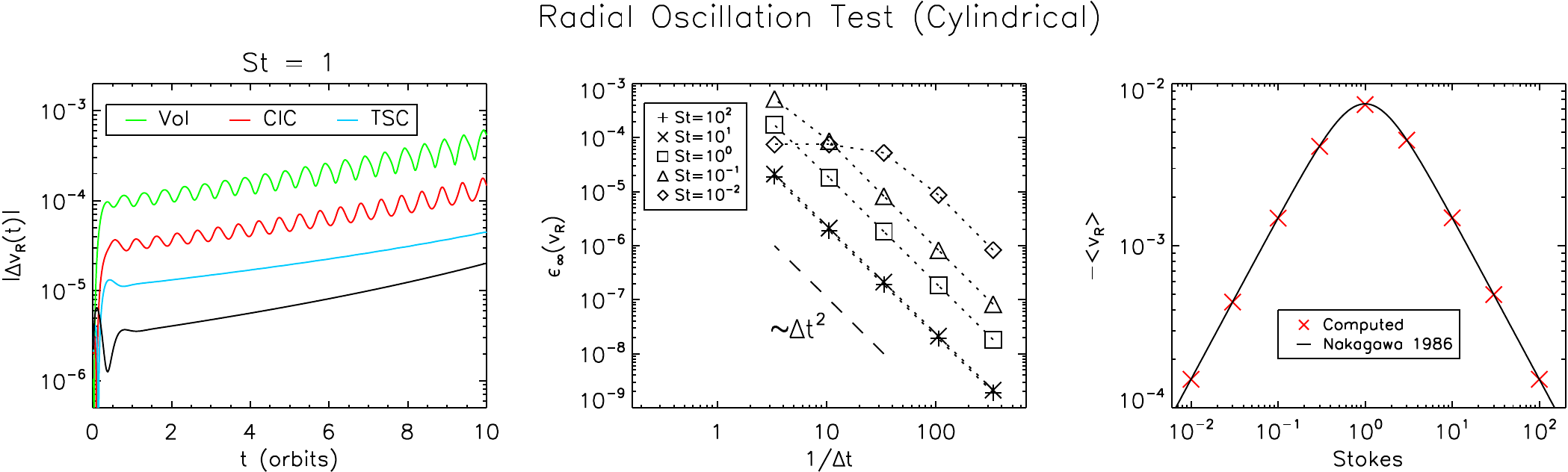}  
  \includegraphics[width=0.75\textwidth]{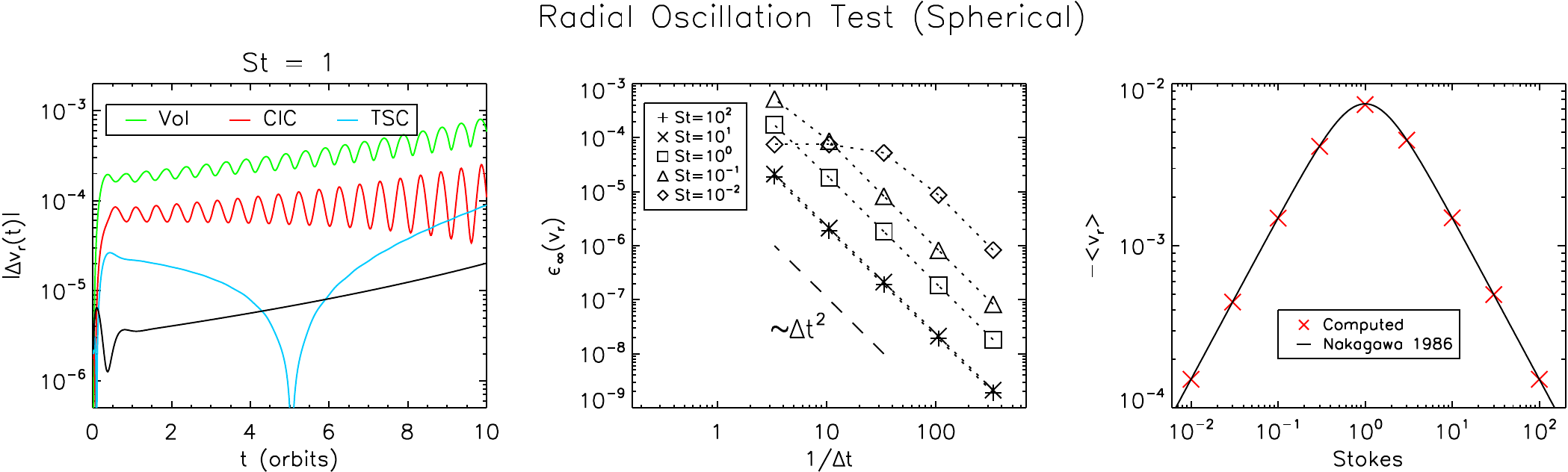}  
  \caption{\footnotesize
           Radial drift test in cylindrical (top row) and spherical (bottom
           row) coordinates.
           \emph{Left panels:} radial velocity errors using different weighting
           factors for particle with Stokes number 1 ($\Delta t = 0.1$ during the
           integration).
           The black line shows the error obtained by assigning gas
           quantities analytically rather than using field weighting during
           particle evolution.
           \emph{Middle panels:} $L_\infty$ radial velocity errors for 
           particles with different Stokes number (reported in the legend).
           \emph{Right panels:} time-averaged radial velocity as a function
           of the particle Stokes number (symbols) and predicted profile \cite{NSH.1986}.
           \label{fig:rdrift}}
\end{figure*}
We now investigate radial drift of grains in a differentially rotating disk with density and angular velocity given by \citep{nelson2013}:
\begin{equation}\label{eq:Nelson_equil}
  \begin{array}{ll}
  \DS \rho &= \DS \rho_0 \left ( \frac{R}{R_0} \right )^p \exp
         \left[\frac{GM}{c^2_s}\left(\frac{1}{r} - \frac{1}{R}\right)\right]\,,
  \\ \noalign{\smallskip}
  \DS \Omega &=\DS
  \Omega_K \left[ (p+q) \left ( \frac{H}{R} \right )^2 + (1 + q)
                   - \frac{qR}{r} \right]^{\frac{1}{2}} \,,
  \end{array}
\end{equation}
where $R$ is the cylindrical radius, $r=\sqrt{R^2 + z^2}$ the spherical radius, $\Omega_K=\sqrt{G M/ R^3}$ the Keplerian angular velocity and the speed of sound is defined by $c_s = H\Omega_K$ with the scale height $H=H_0 \left (R/R_0\right
)^{(q+3)/2}$.
We investigate both axisymmetric cylindrical and spherical geometries allowing particles to evolve in time while gas quantities are fixed during the computations.
In cylindrical coordinates, the computational domain is given by $R\in[0.2,1.6]$ AU, $z\in[-0.2,0.2]$ AU using $64\times 16$ equally spaced zones.
In spherical coordinates, computations are performed in the domain $r\in[0.2,1.8]$ AU, $\theta\in[\pi/2-0.2,\pi/2+0.2]$ using the same resolution.
To represent a typical disk profile we choose $p=-1$ and $q=-0.5$ and $H_0=0.1$ at $R_0=1$.

We place 10 particles in the equatorial plane at $R=1$ with Stokes numbers ${\rm St}=\tau_s\Omega_K = (10^{-2}, 3\times 10^{-2}, 10^{-1}, 0.3, 1, 3,10,30, 10^2, \infty)$ initially moving with Keplerian velocity.
In such a way $v_{p,\phi} - v_{g,\phi}$ is constant over radius.
Particles are evolved for 10 orbits ($t_{\rm stop} = 20\pi$) using the exponential midpoint method and a fixed time step $\Delta t$.
An accurate reference solution is produced by integrating the equations of motion (\ref{eq:dust_cylindrical}) and (\ref{eq:dust_spherical}) with an adaptive step-size $5^{\rm th}$ order Runge-Kutta method.

Results are shown in Fig.~(\ref{fig:rdrift}) for cylindrical (top) and spherical (bottom) coordinates.
A comparison between different weighting schemes is given in the left panels where we plot, as a function of time, the errors in the radial velocity for the particle with $St=1$ with $\Delta t=0.1$.
The time step has been chosen large enough in order to emphasize the error although smaller $\Delta t$ are employed in typical simulations.
Radial volume weighting (Eq. \ref{eq:CIC_vol}) produces the largest errors which are significantly reduced by more than one order of magnitude when switching to TSC weighting (Eq. \ref{eq:TSC_cylindrical} and Eq. \ref{eq:TSC_spherical_r} for cylindrical and spherical geometries, respectively).
As a reference, we also plot the errors obtained by replacing interpolation of gas quantities (field weighting) with direct analytical assignment at the particle position (black lines).

Compared to previous works \cite[see, e.g.,][where similar tests are presented]{Zhu_etal.2014, Stoll_Kley.2016}, we further show a time-resolution study in the middle panels by plotting the maximum errors in the radial velocity for particles with different Stokes numbers.
We choose the time steps as $\Delta t = 3\times 10^{-k/2}$ ($k= 2...6$) in order to cover two decades.
Gas velocity is obtained from direct analytical assignment (no field weighting from fluid to particles) in order to assess only the temporal accuracy of our particle pusher.
For time steps smaller than the particle's stopping time ($\Delta t < \tau_s$), computed solutions converge with second-order rate, as expected.
For ${\rm St} < 10$ errors are roughly proportional to the inverse of the Stokes number, i.e., $\epsilon_{1} \sim {\rm St}^{-1}$.
On the contrary, when the temporal resolution is below the particle's stopping time ($\Delta t > \tau_s$), we found that the error saturates at approximately $\Delta t \gtrsim 10\tau_s$ (see, .e.g, particle with Stokes number $10^{-2}$ in the plot).
This apparently odd behavior should not be surprising as the exponential midpoint method captures the stiff drag term exactly while remaining $2^{\rm nd}$-order accurate when nonlinear terms are present.
In the stiff regime ($\tau_s \ll \Delta t$), in fact, numerical integration is dominated by the linear drag term while the contribution to the error from centrifugal and gravitational terms becomes less important.
As a consequence, numerical integration will be (unexpectedly) more accurate in this regime.

Finally, in the rightmost panels of the same figure we plot the time-averaged radial velocity for different Stokes numbers and compare our results with the predicted radial drift \cite{NSH.1986}, which is maximized for particles with Stokes number equal to unity:
\begin{equation}
  v^{\rm drift} = \frac{\partial\log p}{\partial\log R}
                   \frac{(H/R)^2 v_K}{{\rm St} + {\rm St}^{-1}}
                = -\frac{3}{2}\left(\frac{H_0}{R_0}\right)^2
                  \frac{{\rm St}}{1 + {\rm St}^2} v_K \,,
\end{equation}
where the last equality results from setting $p = 2q = -1$ in Eq. (\ref{eq:Nelson_equil}).
Our results indicate excellent agreement with theoretical predictions.

\subsection{Vertical Oscillations at different Stokes' Numbers}
%

\begin{figure}
  \centering
  \includegraphics[width=0.4\textwidth]{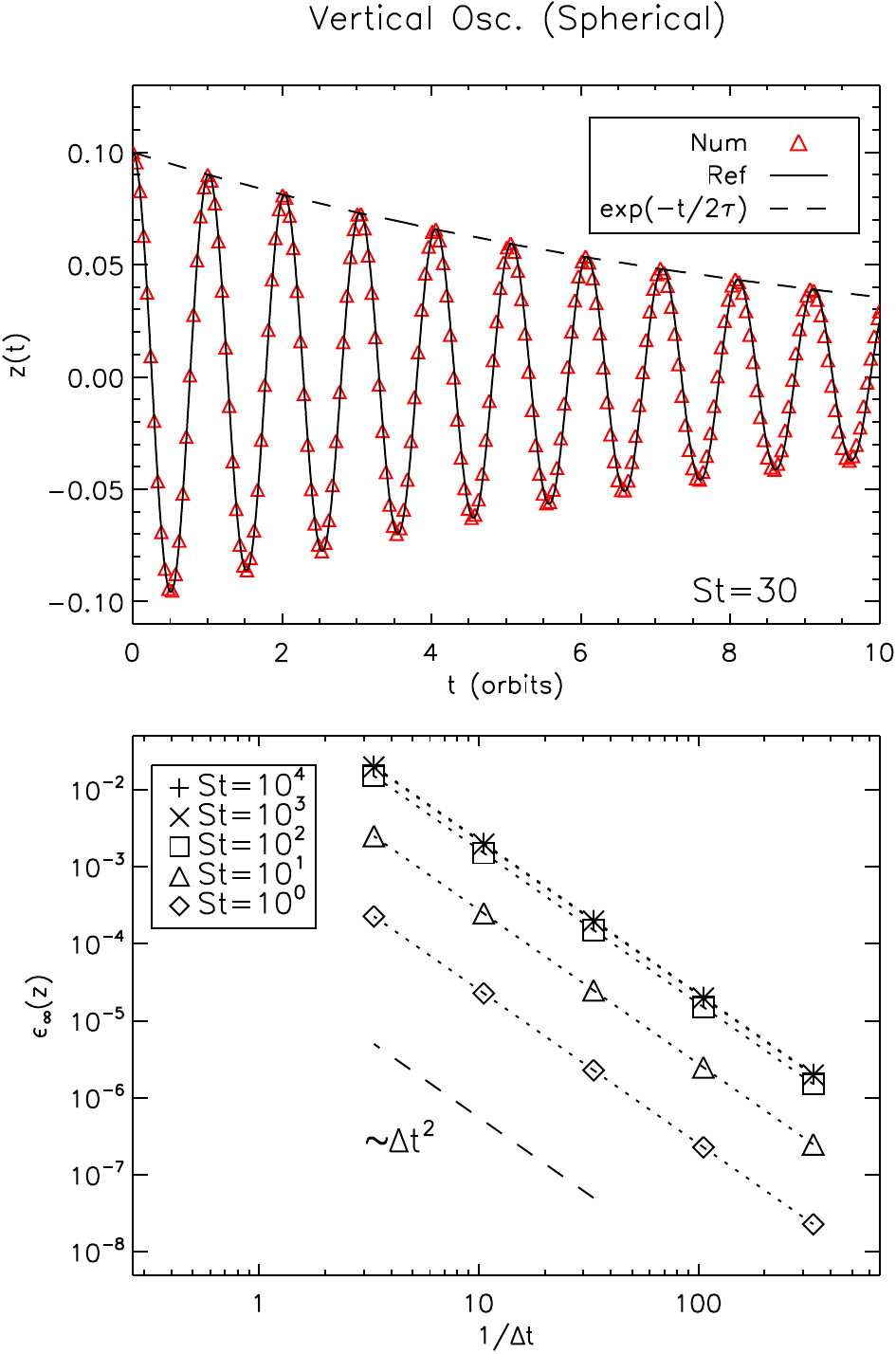}%
  \caption{\footnotesize
           Vertical oscillation test in spherical coordinates.
           \emph{Top panels:} vertical coordinate $z=r\cos\theta$ as a
           function of time.
           Red symbols correspond to numerical integration with $\Delta t = 0.3$
           while the solid black line gives the reference solution.
           \emph{Bottom panel:} convergence study for particles with different
           Stokes number (reported in the legend).
           \label{fig:vdrift_spherical}}
\end{figure}

Dust settling is a crucial process in the formation of planetesimals.
\BLUE
Owing to the presence of a vertical gravity components, grains start to settle towards the disk mid-plane from early on.
The time scale of this process is regulated by the interaction between gas and dust.
\BLACK

In this test, we consider the vertical motion of particles and adopt the same configuration used in the previous one but set $p$ and $q$ to zero which has the advantage that radial drift is negligible.
\BLUE
Ten particles with (constant) Stokes numbers ${\rm St}=\tau_s\Omega_K = (1, 3, 10, 30, 100, 300, 10^3,3\cdot10^3, 10^4, \infty)$ are placed at 1 scale height above the mid-plane ($z=0.1$, $R=1$).
\BLACK
%
%
Particles undergo damped oscillations around the mid-plane with an e-folding time of $\approx 2\tau_s$, while the frequency of oscillations is given by the Keplerian angular velocity.
In the top panel of Fig.~\ref{fig:vdrift_spherical} we plot, for a particle with Stokes number $30$, the vertical position above the mid-plane as a function of time in spherical coordinates (the cylindrical test produces nearly identical results).
The numerical solution, obtained with a fixed time step $\Delta t = 0.3$ (symbols) is shown along with the predicted value (solid line, obtained with a highly resolved $5^{\rm th}$ order adaptive Runge-Kutta method).
Note that the time step exceeds the nominal values used in typical computations, where already with as few as $8$ zones per scale-height one would expect $\Delta t \lesssim 0.1$

In the bottom panel we show a time-resolution study by plotting the $L_\infty$ norm error of $z(t)$ for particles with different Stokes numbers.
Errors scale as $\Delta t^2$ for all particles and are comparable when ${\rm St} \gtrsim 100$ since damping is weak and phase error gives the dominant contribution.
At smaller Stokes number, on the contrary, the solution is strongly damped and phase errors become negligible thus producing lesser errors.

\subsection{Rigidly Rotating Disk with Feedback}
\label{sec:rigid_disk}
%

In the next test problem we assess the accuracy of the fluid-particle feedback term implementation in non-Cartesian geometry.
\BLUE
To this end, we set up a disk with spatially uniform densities $\rho_g(t)$ and $\rho_\D(t)$ (for gas and dust, respectively) embedded in a radial gravitational field linearly increasing with distance, $\vec{g} = (-\omega^2R,\, 0,\, 0)$.
Velocities are assumed to have a similar radial dependence:
\begin{equation}
  \begin{array}{l}
  \vec{v}_g = \big(\alpha(t),\, \Omega_g(t),\, 0\big)R \,,
  \\ \noalign{\smallskip}
  \vec{v}_\D = \big(\beta(t), \, \Omega_\D(t),\, 0\big)R \,,
  \end{array}
\end{equation}
where $R$ is the cylindrical radius, $\alpha(t)R$ and $\beta(t)R$ are the radial velocities while $\Omega_{g}(t)$ and $\Omega_\D(t)$ are the angular rotation velocities for the gas and dust components, respectively.
Separation of variables allows to remove the radial dependence from the gas-dynamical equations reducing them to a simple time-dependent system of ODE:
\begin{equation}\label{eq:rigid_disk_ODEg}
  \begin{array}{l}
    \DS \frac{d\rho_g(t)}{dt} = - (n+1)\rho_g(t)\alpha(t) \,,
    \\ \noalign{\smallskip}
    \DS \frac{d\alpha(t)}{dt} = - \alpha^2(t) + \Omega^2_g(t) - \omega^2 
                              - \frac{\rho_\D(t)}{\rho_g(t)}\frac{\alpha(t) - \beta(t)}{\tau_s}\,,
    \\ \noalign{\smallskip}
    \DS \frac{d\Omega_g(t)}{dt} = - 2\alpha(t)\Omega_g(t)
                                  - \frac{\rho_\D(t)}{\rho_g(t)}
                                    \frac{\Omega_g(t) - \Omega_\D(t)}{\tau_s}\,,
  \end{array}
\end{equation}
which correspond to mass continuity, radial and azimuthal components of the momentum equations.
Here $n=1$ and $n=2$ are used for cylindrical and spherical geometries (the latter is valid only in the equatorial plane, $\theta = \pi/2$).
The first three terms on the right hand side of the radial momentum equations account, respectively, for radial inflow ($-\vec{v}_g\cdot\nabla\vec{v}_g$), centrifugal ($v_\phi^2/R$) and gravitational terms.
Likewise, the dust equations (in the fluid description) take the following form:
\begin{equation}\label{eq:rigid_disk_ODEd}
  \begin{array}{l}
    \DS \frac{d\rho_\D(t)}{dt} = - (n+1)\rho_\D(t)\beta(t)\,,
    \\ \noalign{\smallskip}
    \DS \frac{d\beta(t)}{dt}  = - \beta^2(t)^2 + \Omega^2_\D(t) - \omega^2 
                                + \frac{\alpha(t) - \beta(t)}{\tau_s}\,,
    \\ \noalign{\smallskip}
    \DS \frac{d\Omega_\D(t)}{dt} = - 2\beta(t)\Omega_\D(t)
                                   + \frac{\Omega_g(t) - \Omega_\D(t)}{\tau_s}\,.
  \end{array}
\end{equation}

We set our units so that $\omega = 1$.
In absence of viscous drag, a rotationally-balanced equilibrium is trivially obtained by setting $\alpha(0) = 0$ and $\Omega_g(0) = \Omega_g(0) = \omega$.
Here, however, we consider a time-dependent version of this problem with fluid and dust initially rotating at slightly different orbital velocities,
\begin{equation}
  \Omega_g(0) = \frac{8}{10}\omega,\quad \Omega_\D(0) = \omega \,,
\end{equation}
with density $\rho_g(0) = 1$, $\rho_\D(0) = \rho_g(0)\epsilon = 0.2$.
Radial velocities are initially zero, $\alpha(0) = \beta(0) = 0$ while the particle stopping time is taken to be $\tau_s = 1/\omega = 1$.

The systems of ODEs (\ref{eq:rigid_disk_ODEg}) and (\ref{eq:rigid_disk_ODEd}) are solved simultaneously by means of a high-order Runge-Kutta scheme in order to obtain a reference solution for the six unknowns $\rho_g(t)$, $\rho_\D(t)$, $\Omega_g(t)$, $\Omega_\D(t)$ and $\alpha(t),\beta(t)$.

\BLACK
We perform computations inside the 1D domain $R\in[R_b,R_e]$ with $R_b=0$, $R_e=6$ using a fixed mesh spacing $\Delta R = 6/N_R$ with $N_R=800$.
The $2^{\rm nd}$-order Runge-Kutta scheme with linear reconstruction are used to advance fluid variables with a Courant number $C_a=0.45$.
While fluid quantities can be trivially initialized on the grid, particle assignment requires some additional considerations, \BLUE which we discuss in \S\ref{sec:rigid_disk_particle_assign}.

\BLACK
To avoid dealing with complex boundary conditions, we treat the inner ($R<1$) and outer ($R > 5$) regions as boundary placeholders.
At each time step, we replace the fluid distribution inside these regions by a flat  (for gas density) or linear profiles (for radial and azimuthal velocities) by properly interpolating from the first and last zones of the active domain ($R\in[1,5]$).
Particles are free to move inside the buffer regions so that our results remain valid as long as their radial displacement is less than $1$.
Given the oscillatory character of the solution, this conditions is always respected in practice.
Computations are stopped at $t=2\pi$ (in units of $\Omega^{-1}$).

\BLUE
\subsubsection{Particle Assignment}
\label{sec:rigid_disk_particle_assign}
%
\BLACK

\begin{figure}[!h]
  \centering
  \includegraphics[width=0.4\textwidth]{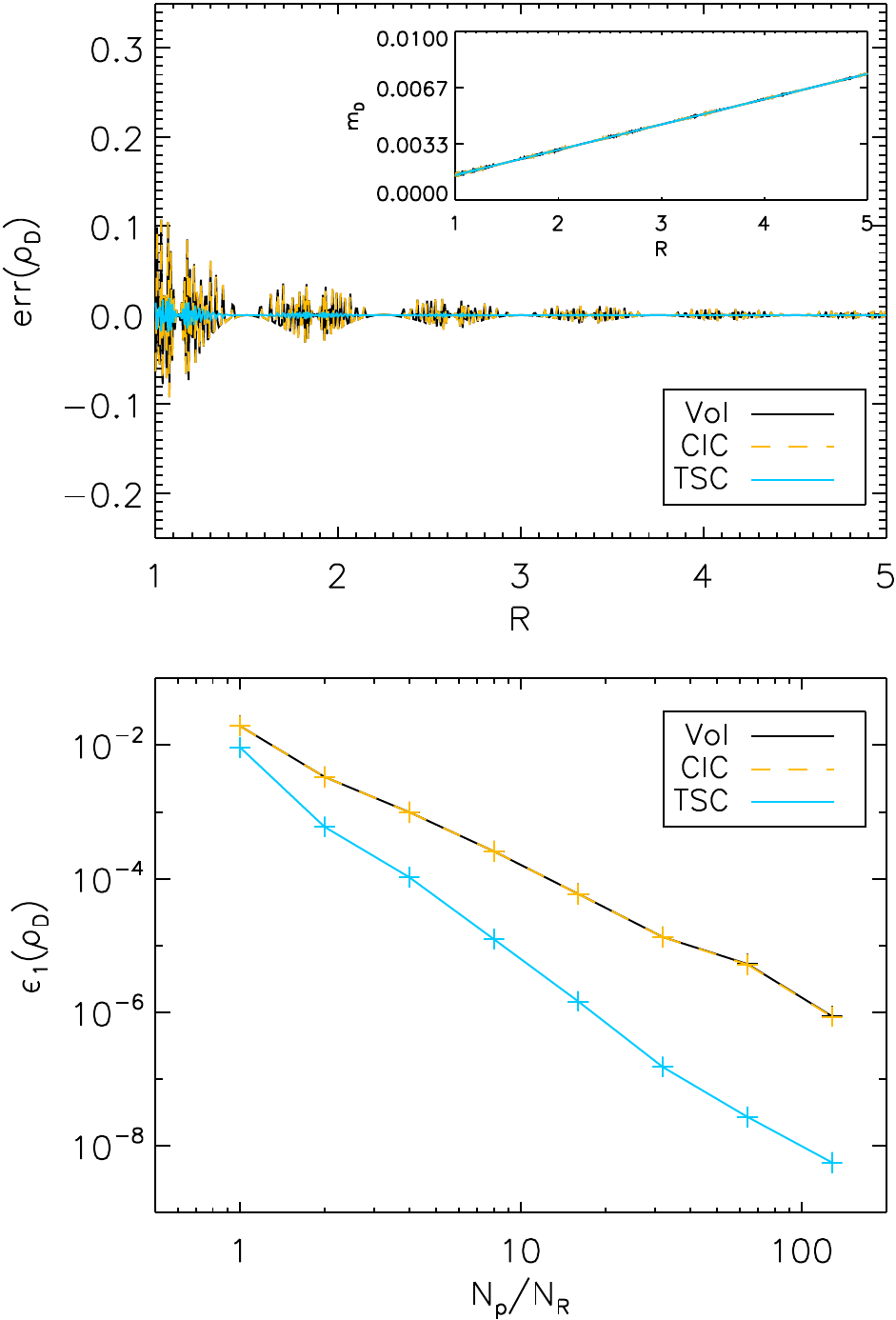}
  \caption{\footnotesize
\BLUE
           \emph{Top:} Relative error ${\rm err}(\rho_\D) = \rho_\D/\epsilon\rho_g-1$           for the uniform dust density
           distribution (at $t=0$) for the rigid disk obtained with
           volume (black solid line), CIC (orange dashed line)
           and TSC (cyan solid line) weighting schemes using $N_p/N_R = 4$.
           \emph{Bottom:} $L_1$ norm error as a function of the number of
           particles (normalized to grid resolution $N_R$).
           \label{fig:distribution}}
\end{figure}

\begin{figure*}
  \centering
  \includegraphics[width=0.9\textwidth]{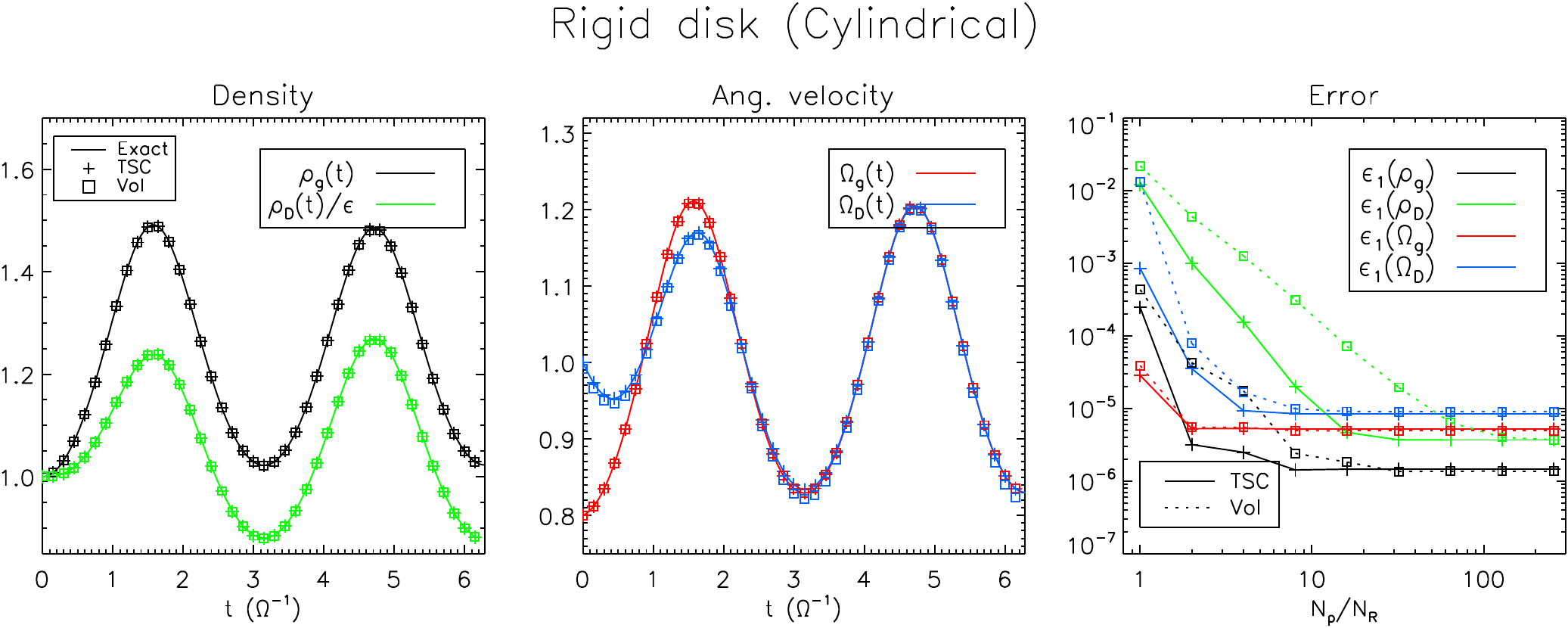}
  \includegraphics[width=0.9\textwidth]{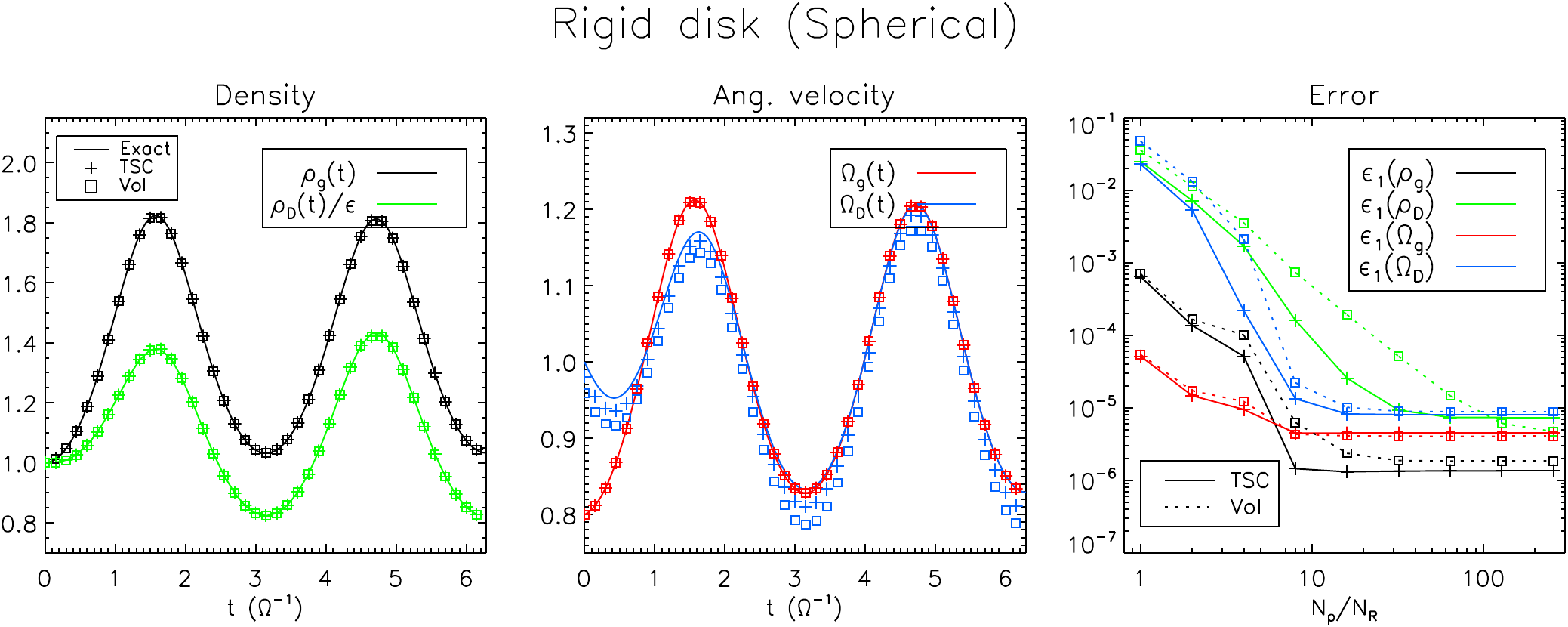}
  \caption{\footnotesize
\BLUE
           Temporal evolution for the rigidly rotating disk in cylindrical
           (top) and spherical (bottom) coordinates.
           Left panel: analytical (solid lines) and numerical solution
           (plus and squares symbols) of density as a function
           of time.
           Red and green curves correspond to $\rho_g(t)$ and
           $\rho_\D(t)\epsilon$.
           Central panel: time evolution of $v_\phi/R=\Omega_g(t)$ (red) and
           $v_{\D,\phi}/R = \Omega_\D(t)$ (blue).
           Right panel: $L_1$ norm error for the different quantities as a function
           of the initial number of particles per volume element.
\BLACK
           \label{fig:rigid_disk}}
\end{figure*}

In order to produce a constant-density distribution, one cannot simply place a constant number of particles in each cell as this would produce a constant-mass distribution $\rho_\D \sim 1/R$ (rather than constant density).
Instead, we use an approach based on the inversion of the the cumulative distribution function \citep{Birsdall_and_Langdon.2004}:
\begin{equation}\label{eq:cumulative_distrib}
   \xi(R) = \frac{\DS\int_{R_b}^{R}   \rho_\D(r) r \,dr}
                 {\DS\int_{R_b}^{R_e} \rho_\D(r) r \,dr}  \,,
\end{equation}
and then equate $\xi(R)$ to a uniform distribution of $N_p$ real numbers in $(0,1)$.
Inverting the previous equation for $R$ yields
\begin{equation}\label{eq:Rxi_constant}
  R(\xi) = \sqrt{\xi (R_e^2 - R_b^2) + R_b^2},
\end{equation}
i.e., constant volumetric distance between particles.
However, since volume elements are nonlinear in the spatial variable for curvilinear coordinates \citep{Verboncoeur.2001}, systematic noise is produced throughout the grid being more severe at smaller radii, \BLUE where fewer particles are located.

To better understand this, let us compute $\Delta\xi(R)\equiv\xi(R+\Delta R) - \xi(R)$, i.e., the spacing of the uniform distribution numbers such that two neighbor particles are one cell away.
From Eq. (\ref{eq:cumulative_distrib}) one finds
\begin{equation}\label{eq:Delta_xi}
  \Delta\xi(R) =  \frac{1}{N_R^2}\frac{R_e - R_b + 2N_RR}{R_e + R_b} \,.
\end{equation}
In order to have at least one particle per cell everywhere in the domain, one has to consider the minimum of $\Delta\xi(R)$ which occurs for $R=R_b$.
Since the $\xi$'s are taken to be equally distributed in $(0,1)$, the total number of particles is given by $N_p\approx 1/\Delta\xi(R_b)$ which, in the limit $R_b\to 0$ gives $N_p\approx N_R^2 = 64\times 10^4$.
If we consider only the active zones ($1\le R \le 5$), this constraint lowers to $N_p = 6N_R^2/(2N_R+4)\approx 2394$ particles.
Since the particles' radial positions will not be evenly spaced, the amount of grid noise depends on the shape function.

The situation is best illustrated in the top panel of Fig. \ref{fig:distribution} showing the relative error in the initial dust density distribution obtained with volume, CIC and TSC weighting schemes with $N_p/N_R=4$ in cylindrical coordinates.
Larger errors are produced by schemes with 2-zone support (volume and CIC scheme) while TSC yields smaller deviations and faster convergence as shown in the bottom panel.
Here we plot the $L_1$ norm error
\begin{equation}\label{eq:rigid_err}
  \epsilon_{1}(\rho_\D)
  = \frac{\sum_i |\rho_{\D,i}-\epsilon\rho_{g,i}|\Delta V_i} 
         {\sum_i \Delta V_i} \,,
\end{equation}
as a function of the total number of particles over radial resolution $N_p/N_R$.
In Eq. (\ref{eq:rigid_err}), $\Delta V_i$ is the cell volume and the summation extends to active zones only ($1\le R\le 5$, see below).

\BLACK

\subsubsection{Results}
%

Fig.~\ref{fig:rigid_disk} shows the temporal evolution of density and azimuthal velocity coefficients, for the gas and dust components, obtained with $N_p = N_R$ (left and middle panels).
Top and bottom panels corresponds to the cylindrical and spherical geometries, respectively.
Numerical solutions are plotted, together with the expected value (solid lines), using the TSC interpolation scheme (plus sign) as well as the traditional volume-weighting interpolation (squares).
After an initial transient ($t \lesssim \tau_s$), fluid and dust components begin to co-rotate with a time-oscillatory pattern.
\BLUE
The $L_1$ norm errors at $t=2\pi$ are computed using Eq. (\ref{eq:rigid_err}) for different values of $N_p/N_R$ and plotted in the right panel of the same figure.
\BLACK
Solid and dashed lines correspond to the TSC and volume interpolation scheme, respectively.
Larger errors are observed in the dust density and azimuthal velocity due to the initial interpolation noise.
While noise is reduced as the number of particles increases, the TSC scheme yields significantly smaller errors indicating better smoothness properties.
As the number of particles is increased, the solution converges to the truncation level of the scheme albeit the error on dust density decreases faster with the TSC scheme.


\subsection{Streaming Instability in Global Disks}
%

\begin{figure*}[!ht]
  \centering
  \includegraphics[width=0.9\textwidth]{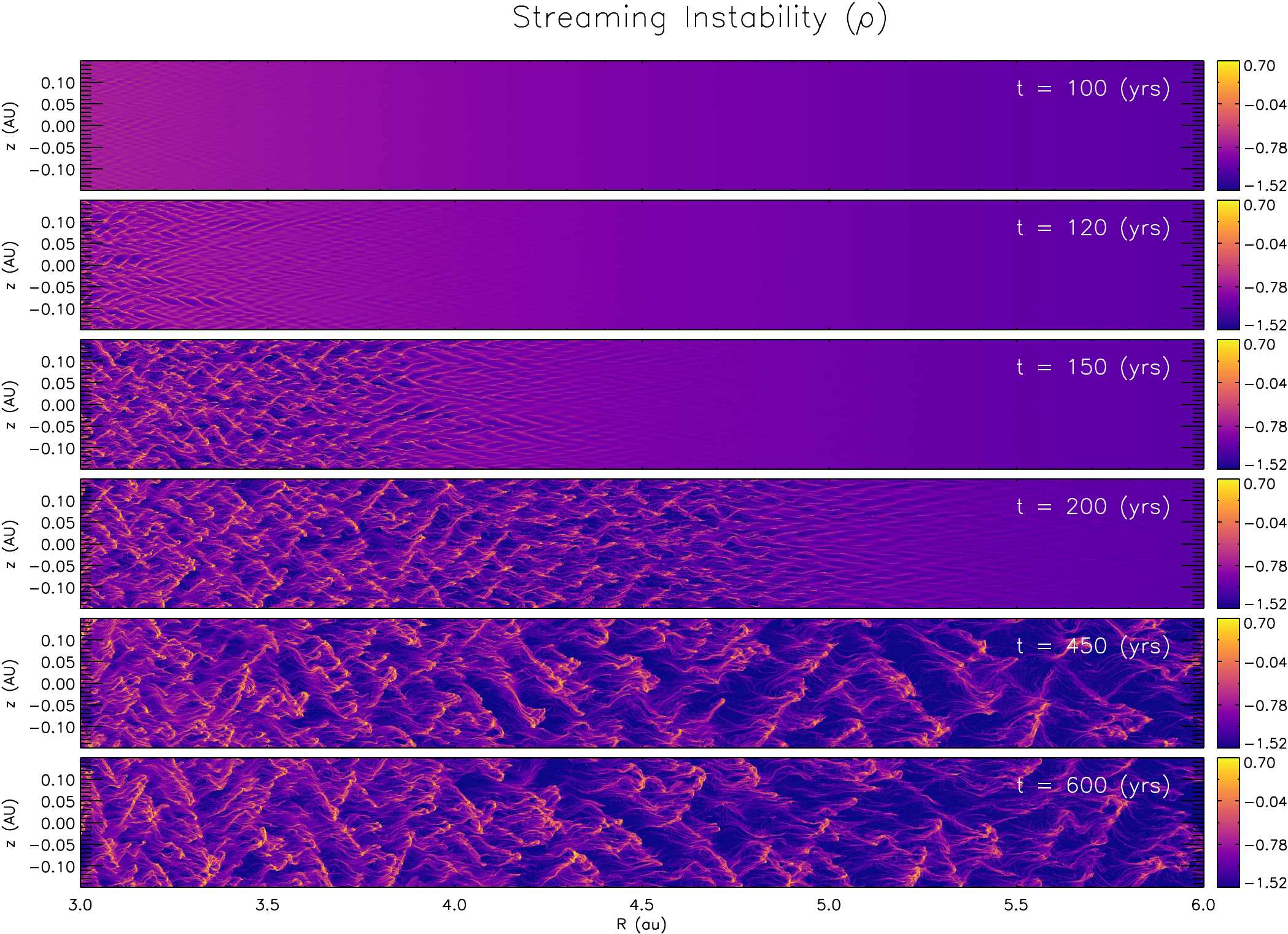}
  \caption{\footnotesize
           Colored maps of the (log of) of dust density for the streaming
           instability test in cylindrical coordinates.
           From top to bottom we show different snapshots
           at $t=100, 200, 450, 600$ (years).
           The resolution is $6144\times 384$ grid zone and only the region 
           $3<R<6$ is shown.
           \label{fig:streaming_global}}
\end{figure*}
As a pilot application, we verify our implementation of the particle-gas hybrid scheme by studying the streaming instability in a non-stratified disk in 2D cylindrical coordinates.
Our setup follows from \cite{Kowalik_etal.2013} and exemplifies a differentially rotating disk in which gas and particle components are separately in equilibrium.

We choose our physical units so that $R_0 = 1\,{\rm AU}$ and $v_0 = \sqrt{GM_\odot/R_0}$ are, respectively, the reference length and velocity while $\rho_0 = \Sigma_0/R_0$ (with $\Sigma_0 = 1700 \,{\rm g/cm}^2$ being the surface density at $R_0$) gives the reference density.
The initial gas equilibrium state is described by Eq. (\ref{eq:Nelson_equil}) with $p = -3/2$ and $q=0$:
\begin{equation}
  \begin{array}{l}
  \DS \rho(R)    = \rho_0  \left(\frac{R_0}{R}\right)^{3/2} \,,
  \\ \noalign{\smallskip}
  \DS \Omega(R)  = \Omega_0\left(\frac{R_0}{R}\right)^{3/2}
                           \sqrt{1 - \frac{3}{2}\frac{c_s^2}{v_0^2}\frac{R}{R_0}} \,,
  \end{array}
\end{equation}
where $\Omega_0 = v_0/R_0$, $c_s = 0.04v_0$ is the isothermal speed of sound corresponding to a constant temperature of $T_0\approx170\, {\rm K}$.
Note that the gas rotates at sub-Keplerian velocity since the radial component of gravity is partially supported by the gas pressure gradient (besides the centrifugal term).
The vertical component gravity is neglected as in \cite{Kowalik_etal.2013}.
Conversely, the dust component is pressureless and the equilibrium is dictated by simple balance between gravitational and centrifugal forces,
\begin{equation}\label{eq:streaming_global_rhoD}
  \rho_\D(R)     = \epsilon\rho(R) \,,\quad
  \Omega_{\D}(R) \equiv \Omega_K(R) = \Omega_0\left(\frac{R_0}{R}\right)^{3/2}  \,,
\end{equation}
where $\epsilon$ is the dust to gas mass ratio.
In the Epstein regime the particle stopping time is prescribed by
\begin{equation}\label{eq:streaming_global_taup}
  \tau_p(R) = \frac{\rho_{\bullet}a}{\rho(R) c_s} \,,
\end{equation}
where $\rho_{\bullet} = 1.6\,{\rm g/cm^3}$ is the solid material density while $a$ (in cm) is the grain size.
We set the grain size to $a=50\,{\rm cm}$ which corresponds to marginally coupled boulders having ${\rm St} = \Omega(R)\tau_p(R)\approx 1.2$.

The computational domain extends over the region $R\in[R_b, R_e]$, $z\in[z_b,z_e]$ and it is uniformly discretized over $N_R\times N_z$ grid zones.
We choose $R_b = 2R_0$, $R_e = 7R_0$ and $z_e = -z_b = 0.15 R_0$.
In order to create the desired particle density distribution given by Eq. (\ref{eq:streaming_global_rhoD}), we invert the cumulative distribution function as in Eq. (\ref{eq:cumulative_distrib}).
This yields the particle radial position:
\begin{equation}
  R(\xi) = -2\xi(\xi-1)\sqrt{R_bR_e} + (R_b + R_e)\xi^2 + R_b(1 - 2\xi) \,,
\end{equation}
while the vertical position is obtained similarly:
\begin{equation}
  z(\zeta) = z_b + \zeta(z_e - z_b) \,.
\end{equation}
In the previous expressions, $\xi$ and $\zeta$ are uniformly distributed numbers in the unit interval.
\BLUE
Since the mass is decreasing with radius, particle sampling is now denser at small radii (as opposed to the previous test case).
In order to have at least one particle per cell, we adopt the same arguments used in \S\ref{sec:rigid_disk_particle_assign} leading to a less restrictive condition, namely, $N_p\approx 1/\Delta\xi(R_e) \approx 1.23/(\sqrt{7+5/N_R}-2.64)$.
\BLACK
We employ $N_{pR}=4N_R$ and $N_{pz}=2N_z$ particles in the $R$ and $z$ directions, respectively, so that $N_p = 8N_{R}N_{z}$ is the total particle number.
The individual particle mass is obtained from the total dust to gas mass ratio:
\begin{equation}
  m_p = \epsilon\frac{M_g}{N_p}
      = \epsilon \frac{2 (z_e-z_b)(\sqrt{R_e} - \sqrt{R_b})}{N_p}
        \rho_0R_0^{3/2} \,,
\end{equation}
where $M_g$ is the total mass of the gas.

Periodic boundary conditions for the fluid and dust grains are imposed at the top and bottom vertical sides of the box.
In the radial direction, fluid boundary values are held fixed to the initial condition and, in addition, we implement a wave killing zone by letting the solution relax to the initial equilibrium values,
\begin{equation}
  Q(\vec{x},t) =   Q_0(\vec{x})
                 + \Big[Q(\vec{x},t) - Q_0(\vec{x})\Big]e^{F(R) \Delta t/T} \,,
\end{equation}
where $Q(\vec{x},t)$ is any fluid variable, $Q_0(\vec{x})$ is the corresponding equilibrium value, $T = 2\pi\sqrt{2}$ and
\begin{equation}
   F(R) = 2 - \tanh\left(\frac{R - R_b}{w}\right)^8
            - \tanh\left(\frac{R - R_e}{w}\right)^8 
\end{equation}
is a tapering function with $w = 0.2R_0$.
Particles are free to move in the buffer regions and are removed from the computational domain once they cross the inner boundary.
At the outer boundary, instead, new particles must be injected as the radial drift empties the wave killing area.
We achieve this by creating new particles (with equilibrium values) whenever the region $R/R_0>6.9$ is depleted of dust.

We perform computations at three different resolutions, $N_R = 1536,\, 3072,\, 6144$ and $N_z = N_R/16$ to have approximately square cells.
Integrations are carried for $\approx 600$ years using the second-order Runge-Kutta scheme (for the fluid) with a Courant number $0.45$ and the exponential midpoint method for particles.
We set the initial dust to mass ratio to $\epsilon=1$ which corresponds to case `BB' in the notations of \cite{JY.2007} and \cite{Kowalik_etal.2013}.

In Fig.~\ref{fig:streaming_global}, we show colored maps of the dust density at $t=100, 200, 450$ and $600$ years of evolution.
The early stages are characterized by the amplification of local over-densities turning into stretched clumpy filaments mainly aligned with the radial direction.
Owing to the differential rotation, the instability proceeds faster at smaller radii and slower as we move away from the center.
After the linear phase, the instability enters in the nonlinear regime and over-dense filaments tend to form larger structure leaning along the diagonal directions.

\begin{figure}
  \centering
  \includegraphics[width=0.45\textwidth]{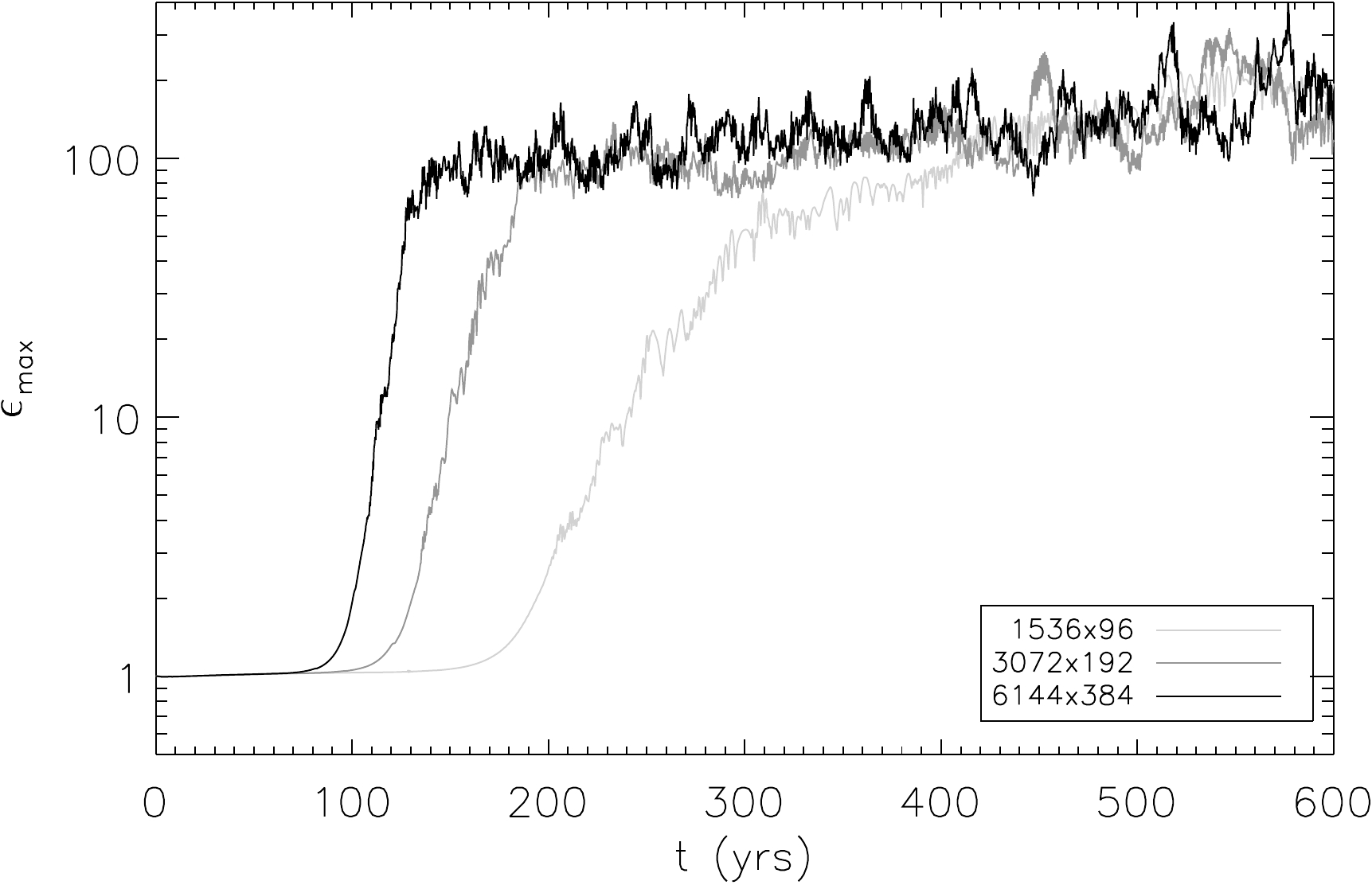}
  \caption{\footnotesize
           Maximum dust-to-gas density ratio over time at different resolutions
           in the streaming instability problem in 2D cylindrical coordinates.
           \label{fig:streaming_global_eps}}
\end{figure}
In Fig.~\ref{fig:streaming_global_eps}, we plot the maximum ratio of $\epsilon=\rho_\D/\rho$, taken in the range $3 \le R/R_0 \le 6$ as a function of time.
As in \cite{Kowalik_etal.2013}, larger resolutions trigger short-wavelength faster growing modes and the instability sets off earlier.
During the nonlinear stages, local over-densities reach about 2 order of magnitude ($\epsilon\approx 100$) the initial value albeit this growth is considerably slower for the low-resolution run.

\begin{figure}
  \centering
  \includegraphics[width=0.45\textwidth]{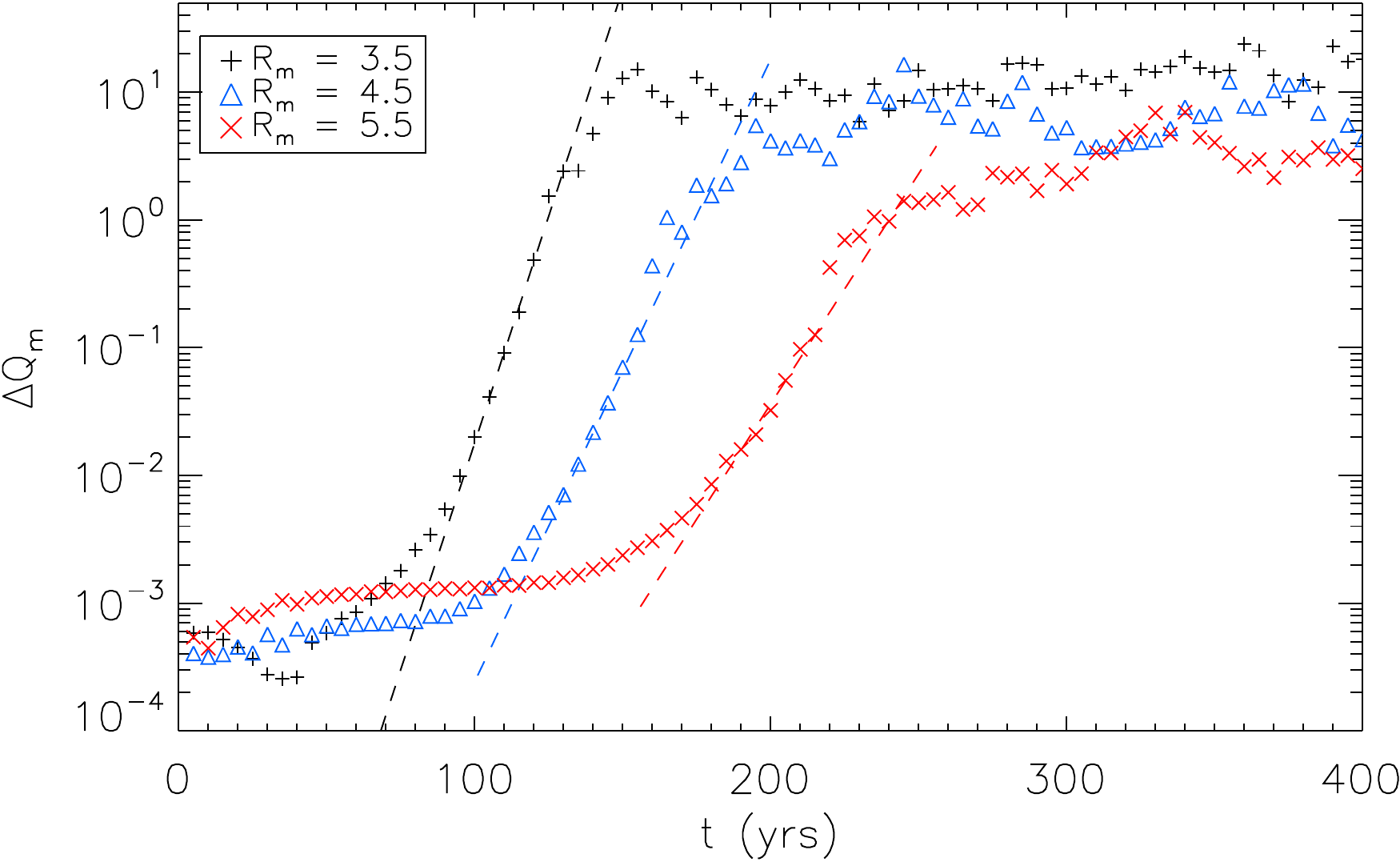}
  \caption{\footnotesize
           Measured growth rates (symbols) and analytical values (dashed lines) 
           for the global streaming instability problem.
           Different colors correspond to different radial patches centered 
           at $R_m$ (see the legend).
           \label{fig:streaming_global_grate}}
\end{figure}

It is instructive to compare the growth of the instability during the linear phase with the values predicted from linear theory.
To this end we have employed the publicly available solver of \cite{Kowalik_etal.2013} to solve the linear dispersion relation in the shearing-sheet approximation for our particular conditions.
%
%
Since $\Omega=\Omega(R)$ is not constant in a differentially rotating disk, we have computed the growth rates at different locations in the domain, namely, for narrow radial patches centered at $R_m=\{3.5, 4.5, 5.5\}R_0$ having width $\Delta w = 0.15 R_0$.
From our initial conditions we determine the input parameters to the dispersion relation as $\epsilon=1$, $\Omega_m = \Omega_0R_m^{-3/2}$, $\eta_m = 3/4 c_s^2/v_{K,m}^2$ while $\tau_p(R_m)$ computed from Eq. (\ref{eq:streaming_global_taup}).
The peak growth rate is found to be $s_m/\Omega_m \approx 0.17$, approximately independent on the radial distance.
For each patch $m$ we evaluate the quantity
\begin{equation}
  \Delta Q_m(t) = \max_m\left|\frac{\rho_\D(R,z,t) - \rho_\D(R, z, 0)}{\rho_0}\right| \,,
\end{equation}
where the maximum is taken between $R_m-\Delta w/2$ and $R_m+\Delta w/2$.
Plots of $\Delta Q_m$ are shown on a log scale in Fig.~\ref{fig:streaming_global_grate} at the highest resolution together with the analytical value which, neglecting sub-Keplerian motion, is proportional to $\sim\exp(0.17\Omega_m t)$.
Despite the simplified comparison, our results indicate that the linear stage of the instability proceeds at the predicted analytical rate.
A more quantitative analysis is obtained by computing the growth rates from the linear fit of $\log\Delta Q_m(t) \approx s_mt + b$ over an interval of $\approx 7$ revolutions centered around the middle of the growth curve.
Computed values of $s_m/\Omega_m$ are listed in Table \ref{tab:streaming_instability_grate} for different radii and for the chosen grid resolutions.
Although the measured growth rates are subject to some uncertainty (depending on the time window) the best agreement is achieved at the largest resolution while halving the number of grid cells progressively underestimates the growth rate by $\sim 11\%$ and $\sim 40\%$. 

\begin{table}
\begin{center}
\caption{\footnotesize Normalized growth rates $s_m/\Omega_m$ obtained
 at different radii and for the three grid resolutions in the global
 streaming instability test.
 The (approximate) expected value is $\sim 0.17$.
         \label{tab:streaming_instability_grate}}
\begin{tabular}{cccc}
\tableline \noalign{\smallskip}
          & \multicolumn{3}{c}{Grid resolution} \\
            \cline{2-4} \\
   $R_m/R_0$  & $1536 \times 96$   & $3072 \times 192$ & $6144 \times 384$  \\
   \tableline \\
3.5  &  0.081  &  0.115  &  0.162    \\ \noalign{\smallskip}
4.5  &  0.100  &  0.152  &  0.171    \\ \noalign{\smallskip}
5.5  &  0.116  &  0.144  &  0.165    \\ \noalign{\smallskip}
\hline
\end{tabular}
\end{center}
\end{table}

\section{Summary}
\label{sec:summary}
%
%
%
%

A gas-particle hybrid numerical method to simulate the mutual interaction between gas and dust grains has been the subject of this work.
The gas component is treated using a fluid description and the numerical solution leans on traditional finite-volume methods. 
On the contrary, dust is modeled as an ensamble of solid particles treated by particle-in-cell techniques.
In the Epstein regime gas and dust components are coupled via an aerodynamic drag force that is linear in the relative velocity and the proposed framework has been investigated in the context of dust dynamics in protoplanetary accretion discs. 
Our scheme is able to model the large-scale dynamics of gas and dust including back-reaction, allowing relevant instabilities to be well resolved while, at the same time, being capable of tracking the kinematic and thermal histories of individual grains. 
The method, implemented as a new particle module in the PLUTO code, delivers a number of innovative aspects with respect to previously presented schemes.

In order to handle the stiffness induced by the drag term in the particle equations of motion, a novel time-reversible exponential integrator has been presented.
This method belongs to a class of ODE solvers that combines geometric and exponential integrators allowing stable and reliable computations for arbitrary  particle stopping time within the same time-marching algorithm.
Exponential methods are particularly suited for systems of ODE with a vector field separated into a linear term (typically introducing stiffness) and a nonlinear term.
The computational cost and complexity of the method are comparable to the classical (i.e. non-exponential) methods.

Our numerical method has been formulated not only for Cartesian geometry but it has also been extended to cylindrical and spherical coordinates.
Adaptation to curvilinear geometry required changes in the particle pusher as well as in the particle-to-grid traditional weighting schemes.
For the former, the exponential midpoint method has been extended to cylindrical and spherical geometries using angular-momentum conserving form.
For the latter, we have derived second- and third-order weighting schemes that improve over the traditional volume-weighting interpolation.

An extensive suite of numerical benchmarks has been considered to assess the validity of our implementations in different geometries and under diverse particle conditions.
Numerical results yielded stable computations and excellent agreements with analytic or reference solutions.
The streaming instability has been successfully reproduced in both local and global disk models, providing an excellent verification test for particle-gas dynamics with back-reaction. 
Compared to the Cartesian case, computations in cylindrical geometry required a larger average number of particles per cell (typically a factor $\gtrsim 8$) motivated, on the one hand, by the necessity of reproducing radially-varying profiles and, on the other, by the requirement of lessening grid noise at small radii.

\BLUE
In spite of the results established with the presented numerical method, a number of issues still require further investigation.
Particle to mesh weighting, for instance, still suffers from large error near the coordinate origin and can probably be mitigated by the employment spline functions extending over more than 3 zones.
Another concern comes from the potential stiffness that may arise during the fluid integration step, which may be ameliorated by the employment of more stable time-marching scheme.
Some of these issues will be addressed in forthcoming work.
\BLACK

\BLUE
\acknowledgments
We acknowledge the CINECA award under the ISCRA initiative, for the availability of high performance computing resources and support.
Our work has been partially supported by the Prin MIUR grants 2015L5EE2Y.
MF acknowledge funding from the European Research Council (ERC) under the European Union's Horizon 2020 research and innovation programme (grant agreement No. 757957).
BV acknowledges the funding from the Max Planck India Partner group award.
We are also grateful to the anonymous referee for providing us useful insights while improving the quality of this work.
\BLACK

\bibliography{paper}
\bibliographystyle{aasjournal}

\appendix

\section{CTU Hydro Integrator}
\label{app:CTU}
%
%

Here we describe the details of the CTU-time stepping algorithm used during the integration of the hydrodynamics variables.
The solution procedure is very similar to the implementation described in Paper I to which we refer for completeness.
Using the same notation we use $U=(\rho,\, \rho\vec{v}_g)$ and $V=(\rho,\, \vec{v}_g)$ to denote the arrays of conserved variables and primitive variables.
\begin{enumerate}
\item
Compute the drag acceleration term using Eq. (\ref{eq:Delta_vp}) and (\ref{eq:f_D});

\item
For each direction, compute normal predictors in primitive variable $V^*_{\indx,\pm}$(see Eq. 102 and 103 of paper I), and correct velocities for the drag term.
For the $x$-direction, e.g., 
\begin{equation}
  \vec{v}^*_{g,\indx,\pm} \leftarrow \vec{v}^*_{g,\indx,\pm}
    - \frac{\Delta t^n}{2\rho^n_\indx}\av{\vec{f}^n_{\D}}_\indx\cdot\hvec{e}_x
\end{equation}
The normal predictors also includes the source terms (\ref{eq:standard_SB_g}) or (\ref{eq:FARGO_SB_g}) at time level $n$ in the Shearing Box equations model.

\item
Solve Riemann problems between normal predictors to get the fluxes $\F^*_{\indx+\HALF\hvec{e}_d}$ and obtain cell-centered conservative variables at the intermediate time level:
\begin{equation}
  U^{n+\HALF}_\indx = U^{n}_\indx + \frac{\Delta t^n}{2}
                      \left(\sum_d{\cal L}^*_\indx + S^n_{\D,\indx}\right)
\end{equation}
where
\begin{equation}\label{eq:CTU_Lop}
  {\cal L}^*_{d,\indx} =
   - \frac{ \F^*_{\indx+\HALF\hvec{e}_d} - \F^*_{\indx-\HALF\hvec{e}_d} }
          { \Delta x_d }
   + S^n_{g,\indx}
\end{equation}
and $S^n_{g,\indx}$ and $S^n_{\D,\indx}$ have been defined in Section \ref{sec:hydro_integrator}.
This is equivalent to Eq. (108) of Paper I.

\item
Evolve particles by a full step using one of the algorithms described in Section \ref{sec:particle_integrator} and compute feedback source term using Eq. (\ref{eq:dust_feedback_final}).
This yields $S^{n+\HALF}_{\D,\indx}$.

\item
Compute transverse predictor by correcting normal predictors with the transverse flux difference contribution.
For the $x$-states this amounts to
\begin{equation}\label{eq:CTU_cc_states}
  U^{n+\HALF}_{\indx, \pm} = U^{*}_{\indx, \pm} + \sum_{d\ne x}{\cal L}^*_{d,\indx}
\end{equation}

\item
Solve Riemann problem with the corner-coupled states to obtain the fluxes $\F^{n+\HALF}_{\indx+\HALF\hvec{e}_d}$ (see Eq. 112 of Paper I) and update cell-centered conservative variables,
\begin{equation}
  U^{n+1}_\indx = U^{n}_\indx + \Delta t^n
                  \left(\sum_d{\cal L}^{n+\HALF}_{d,\indx}
                        + S_{\D,\indx}^{n+\HALF}\right)
\end{equation}
where ${\cal L}^{n+\HALF}_{d,\indx}$ is computed as in Eq. (\ref{eq:CTU_Lop}) using fluxes and source terms available at the half time level $t^{n+\HALF}$. 

\end{enumerate}

\section{Exponential Midpoint Integrator in Different Systems of Coordinates}
\label{app:exp_midpoint}
%
%

We now show how the exponential midpoint method, given by Eq. (\ref{eq:exp_midpoint}), can be employed in the context of the shearingbox equations (Eqns \ref{eq:standard_SB_p} and \ref{eq:FARGO_SB_p}) and extended to the case of cylindrical and spherical geometries, described by Eq. (\ref{eq:dust_cylindrical}) and (\ref{eq:dust_spherical}), respectively.

\subsection{Shearingbox equations}
\label{app:exp_midpoint_SB}
%

For the standard shearingbox equations without orbital scheme (Eq. \ref{eq:standard_SB_p}), the nonlinear term in Eq. (\ref{eq:exp_midpoint}) evaluates to 
\begin{equation}
   \vec{G}(\vec{x},\vec{v},\vec{v}_g) = \left[\begin{array}{c}
            2\Omega(v_y + q\Omega x) \\ \noalign{\medskip}
           - 2v_x\Omega               \\ \noalign{\medskip}
           - \Omega^2z
   \end{array}\right]
    + \frac{\vec{v}_g}{\tau_s}
\end{equation}
Plugging the previous expression into the kick step of Eq. (\ref{eq:exp_midpoint}) leads to a simple linear system in the velocities at the next time level $\vec{v}_p^{n+1}$:
\begin{equation}\label{eq:standard_SB_exp_midpoint}
   \tens{R}\vec{v}^{n+1}
     =  
    e^{-\Delta t/\tau_s}\vec{v}^n
   + h_1 \vec{G}\left(\vec{x}^{n+\HALF},\,
                      \frac{\vec{v}^n}{2},\,
                      \vec{v}_g^{n+\HALF}\right)
\end{equation}
where $\vec{v}^{n+\HALF}_g$ and $\vec{x}^{n+\HALF}$ are, respectively, the gas velocity and particle position evaluated at the half time step $t^{n+\HALF}$ \BLUE while $h_1$ is the exponential propagator given by Eq. (\ref{eq:h1}). \BLACK
The matrix $\tens{R}$ can be easily inverted yielding
\begin{equation}
  \tens{R}^{-1} = \frac{1}{1+\Omega_1^2}
  \left(\begin{array}{ccc}
    1        &  \Omega_1   &  0   \\ \noalign{\medskip}
  -\Omega_1  &     1       &  0   \\ \noalign{\medskip}
     0       &     0       & 1 + \Omega_1^2
  \end{array}\right)
\end{equation}
%
where $\Omega_1 = \Omega h_1$.

When used in conjunction with the orbital advection scheme Eq. (\ref{eq:FARGO_SB_p}) the scheme modifies to 
\begin{equation}
   \vec{G}(\vec{x},\vec{v}',\vec{v}'_g)  = \left[\begin{array}{c}
                  2\Omega v'_y\\ \noalign{\medskip}
                   v'_x\Omega (q-2)  \\ \noalign{\medskip}
                 - \Omega^2z\end{array}\right] + \frac{\vec{v}'_g}{\tau_s}
\end{equation}
Plugging the previous expression into Eq. (\ref{eq:exp_midpoint}) and solving with respect to $(\vec{v}'_p)^{n+1}$ yields, likewise,
\begin{equation}\label{eq:FARGO_SB_exp_midpoint}
  \tens{R}(\vec{v}')^{n+1}
   =
       e^{-\Delta t/\tau_s}(\vec{v}')^n 
     + h_1  \vec{G}\left(\vec{x}^{n+\HALF},\,
                         \frac{(\vec{v}')^n}{2},\,
                         (\vec{v}')_g^{n+\HALF}\right)
\end{equation}
with
\begin{equation}
  \tens{R}^{-1} = \frac{1}{1+\Omega_1\Omega_2}
  \left(\begin{array}{ccc}
    1        &  \Omega_1   &  0   \\ \noalign{\medskip}
  -\Omega_2  &     1       &  0   \\ \noalign{\medskip}
     0       &     0       & 1 + \Omega_1\Omega_2
  \end{array}\right)
\end{equation}
where $\Omega_1 = \Omega h_1$, $\Omega_2 = \Omega_1 (1-q/2)$.

\subsection{Cylindrical and spherical coordinates}
\label{app:exp_midpoint_curv}
%

\BLUE
We now report the explicit expressions for the exponential midpoint integrator  in cylindrical and spherical geometries.
Our derivation starts from the St\"ormer-Verlet method (Eq. \ref{eq:stormer_verlet}) properly modified during the kick step (Eq. \ref{eq:kick_modified}) to account for the viscous drag term.
The (specific) Hamiltonian for the conservative system is written as
\begin{equation}
  {\cal H}(\vec{p},\vec{q}) = \left\{\begin{array}{ll}
  \DS \frac{1}{2} \left(v_R^2 + \frac{l^2_\phi}{R^2} + v_z^2\right) 
       + \Phi (R, \phi, z)  & \quad {\rm (Cylindrical)} \,,
  \\ \noalign{\smallskip}
  \DS \frac{1}{2} \left(v_r^2 + \frac{l^2_\theta}{r^2}
                              + \frac{l^2_\phi}{r^2\sin^2\theta}\right) 
       + \Phi (r, \theta, \phi)  & \quad {\rm (Spherical)} \,.
  \end{array}\right.
\end{equation}
where $\Phi(\vec{q})$ is the potential energy so that external forces are written as $\vec{g} = -\nabla_\vec{q}\Phi$.

\subsubsection{Cylindrical Coordinates}
\label{app:EM_cylindrical}
%
In cylindrical geometry we evolve the spatial coordinates $\vec{q}=(R,\,\phi,\, z)$ and the momentum-like coordinates $\vec{p} = (v_R,\, l_\phi,\, vz)$ where $l_\phi = R v_\phi$ is the particle angular momentum.
The standard update scheme can be summarized as follows.

\begin{itemize}
\item
Drift by $\Delta t/2$: the particle coordinates $(R,\,\phi,\,z)$ are updated by half a time step:
\begin{equation}\label{eq:drift1_cylindrical}
  \left\{\begin{array}{lcl}
  R^{n+\HALF}      & = & \DS R^n + \frac{\Delta t}{2} v^n_R \,,
  \\ \noalign{\smallskip}
  \phi^{n+\HALF}   & = & \DS  \phi^n
                            + \frac{\Delta t}{2}\frac{l^n_\phi}{(R^2)^{n+\HALF}} \,,
  \\ \noalign{\smallskip}
  z^{n+\HALF}      & = & \DS z^n + \frac{\Delta t}{2} v^n_z \,,
  \end{array}\right.
\end{equation}

\item
Kick by $\Delta t$: update $(v_R,\, l_\phi,\, v_z)$ in reversed order for a full time step:
\begin{equation}\label{eq:kick_cylindrical}
  \left\{\begin{array}{lcl}
  v_z^{n+1}    &=&\DS    e^{-\Delta t/\tau_s}v_z^n
                       + h_1\left(g_z + \frac{v_{g,z}}{\tau} \right)^{n+\HALF} \,,
  \\ \noalign{\smallskip}
  l^{n+1}_\phi &=&\DS   e^{-\Delta t/\tau_s}l_\phi^n
                        + h_1\left(  g_\phi R
                        + \frac{l_{g,\phi}}{\tau_s}\right)^{n+\HALF} \,,
  \\ \noalign{\smallskip}
  v_R^{n+1}    &=&\DS e^{-\Delta t/\tau_s}v_R^n + h_1\left(
                      \frac{l_\phi^2}{R^3}
                     + g_R
                     + \frac{v_{g,R}}{\tau_s}\right)^{n+\HALF} \,,
  \end{array}\right.
\end{equation}
where $h_1$ is the propagator (Eq. \ref{eq:h1}) while, in the last expression,
\begin{equation}
  \left(l_\phi^2\right)^{n+\HALF} = \frac{(l_\phi^2)^n + (l_\phi^2)^{n+1}}{2} \,
\end{equation}
comes from averaging the two Hamiltonians at level $n$ and $n+1$.

\item
Drift by $\Delta t/2$: advance particle position by another half a step:
\begin{equation} \label{eq:drift2_cylindrical}
  \left\{\begin{array}{lcl}
  R^{n+1}      & = & \DS R^{n+\HALF} + \frac{\Delta t}{2} v^{n+1}_R \,,
  \\ \noalign{\smallskip}
  \phi^{n+1}   & = & \DS \phi^{n+\HALF}
                       + \frac{\Delta t}{2}\frac{l^{n+1}_\phi}{(R^2)^{n+\HALF}} \,,
  \\ \noalign{\smallskip}
  z^{n+1}      & = & \DS z^{n+\HALF} + \frac{\Delta t}{2} v^{n+1}_z \,.
  \end{array}\right.
\end{equation}
\end{itemize}
In absence of viscous drag, the method reduces to the classical St\"ormer-Verlet scheme and it is similar to the method outlined in the Appendix of \cite{Zhu_etal.2014}.

\subsubsection{Spherical Coordinates}
\label{app:EM_spherical}
%
In spherical coordinates we adopt a similar procedure to evolve the coordinates $\vec{q} = (r,\,\theta,\,\phi)$ and $\vec{p} = (v_r,\, l_\theta,\, l_\phi)$.

\begin{itemize}

\item
Drift by $\Delta t/2$: particle coordinates are first advanced by half a time step,
\begin{equation}
  \left\{\begin{array}{lcl}
  r^{n+\HALF}      & = & \DS r^n + \frac{\Delta t}{2} v^n_r \,,
  \\ \noalign{\smallskip}
  \theta^{n+\HALF} & = &\DS \theta^n
                        + \frac{\Delta t}{2}\frac{l^n_\theta}{(r^2)^{n+\HALF}} \,,
  \\ \noalign{\smallskip}
  \phi^{n+\HALF}   & = & \DS \phi^n
                         + \frac{\Delta t}{2}
                           \frac{l^n_\phi}{(r^2\sin^2\theta)^{n+\HALF}} \,.
  \end{array}\right.
\end{equation}

\item
Kick by $\Delta t$: velocity is updated by a full time step in reversed order:

\begin{equation}
  \left\{\begin{array}{lcl}
  l^{n+1}_\phi &=& \DS  e^{-\Delta t/\tau_s}l_\phi^n
                   + h_1\left(  g_\phi r\sin\theta
                           + \frac{l_{g,\phi}}{\tau_s}\right)^{n+\HALF} \,,
  \\ \noalign{\smallskip}
  l^{n+1}_\theta &=& \DS  e^{-\Delta t/\tau_s}l_\theta^n
                   + h_1\left( \frac{l_\phi^2\cos\theta}{r^2\sin^3\theta}
                           + rg_\theta
                           + \frac{l_{g,\theta}}{\tau_s} \right)^{n+\HALF} \,,
  \\ \noalign{\smallskip}
  v_r^{n+1} &=& \DS e^{-\Delta t/\tau_s}v_r^n + h_1\left(
                \frac{l_\theta^2}{r^3}
               + \frac{l_\phi^2}{r^3\sin^2\theta}
               + g_r + \frac{v_{g,r}}{\tau_s}\right)^{n+\HALF} \,,
  \end{array}\right.
\end{equation}
where $h_1$ is the propagator (Eq. \ref{eq:h1}).
Note that, owing to the non-linearity of the right hand sides, $l_\theta$ is updated \emph{after} $l_\phi$ has reached level $n+1$ and, similarly, $v_r$ \emph{after} $l_\theta$ has reached the next time level.
Similarly to the cylindrical case, $l_\theta^2$ and $l_\phi^2$ at the midpoint level are obtained by averaging the same quantities at the base and next time level:
\begin{equation}
  \left(l_\phi^2\right)^{n+\HALF} = \frac{(l_\phi^2)^n   + (l_\phi^2)^{n+1}}{2}
  \,,\quad
  \left(l_\phi^2\right)^{n+\HALF} = \frac{(l_\theta^2)^n + (l_\theta^2)^{n+1}}{2}\,.
\end{equation}

\item
Drift by $\Delta t/2$: particle coordinates are advanced to the next time level by another half step:

\begin{equation}
  \left\{\begin{array}{lcl}
  r^{n+1}      & = & \DS r^{n+\HALF} + \frac{\Delta t}{2}v^{n+1}_r  \,,
  \\ \noalign{\smallskip}
  \theta^{n+1} & = & \DS \theta^{n+\HALF}
                     + \frac{\Delta t}{2}\frac{l^{n+1}_\theta}{(r^2)^{n+\HALF}} \,,
  \\ \noalign{\smallskip}
  \phi^{n+1} & = & \DS \phi^{n+\HALF}
                      + \frac{\Delta t}{2}\frac{l^{n+1}_\phi}{(r^2\sin^2\theta)^{n+\HALF}} \,.
  \end{array}\right.
\end{equation}

\end{itemize}

In absence of viscous drag, the method reduces to the classical St\"ormer-Verlet scheme.

\BLACK

\section{Weighting factors in Spherical Coordinates}
\label{app:weights_spherical}
%

Weight functions in spherical coordinates can be recovered using Eq. (\ref{eq:shape_curv}) with $(x_1, x_2, x_3) = (r,\theta,\phi)$ while $\Delta V_p = r_p^2\sin(\theta_p)\,\Delta r\,\Delta\theta\,\Delta\phi$.
Integrating the shape function with $m=0$ over the cell volume, Eq. (\ref{eq:weight}), yields the cloud in cell (CIC) weighting scheme for the radial direction:
\begin{equation}\label{eq:CIC_spherical_r}
  \begin{array}{ll}
  W_i
  &=\DS \frac{3(\delta + 2\nu_i)^2 + (|\delta| - 1)^2}{12(\delta + \nu_i)^2+1}
        \;(1 - |\delta|)
  \\ \noalign{\smallskip}
  W_{i\pm1} &=\DS \frac{(2\delta + 3\nu_i \pm 3/2)^2 + 3(\nu_i \pm 1/2)^2}
                       {12(\delta + \nu_i)^2+1}
                  \;\frac{|\delta|\pm\delta}{2}
  \end{array}                                 
\end{equation}
Using the subsequent spline ($m=1$) we obtain the triangular shape cloud (TSC) weight factors for the radial direction,
\begin{equation}\label{eq:TSC_spherical_r}
  \begin{array}{ll}
  W_i
  &=\DS \frac{ \left[(\delta + 2\nu_i)^2 + 2\nu_i^2 + 3/4\right](3/4-\delta^2) - 1/4}
             {6(\delta + \nu_i)^2 + 1}
  \\ \noalign{\smallskip}
  W_{i\pm1} &=\DS \frac{\delta^2 + (4\nu_i \pm 3)\delta + 6\nu_i^2 \pm 8\nu_i + 11/4}
                        {6(\delta + \nu_i)^2 +1}
                   \;\frac{1}{2}\left(\frac{1}{2}\pm\delta\right)^2     
  \end{array}                                 
\end{equation}

For the $\theta$ direction, we report here the CIC scheme only,
\begin{equation}\label{eq:CIC_spherical_theta}
  \begin{array}{ll}
    W_i & = \DS \left\{\begin{array}{ll}
            \DS\frac{\cos(x_{i-\HALF} + \delta) - \cos(x_{i+\HALF})}
                    {\cos(x_{i-\HALF} + \delta) - \cos(x_{i+\HALF} + \delta)}
          & \quad {\rm if}\quad \delta > 0
          \\ \noalign{\smallskip}
            \DS \frac{\cos(x_{i-\HALF}) - \cos(x_{i+\HALF} + \delta)}
                     {\cos(x_{i-\HALF} + \delta) - \cos(x_{i+\HALF} + \delta)}
          & \quad {\rm if}\quad \delta < 0
        \end{array}\right.
    \\ \noalign{\smallskip}
    W_{i\pm1} & =\DS \frac{\pm\cos(x_{i\pm\HALF}) - \cos(x_{i\pm\HALF} + \delta)}
                          {\cos(x_{i-\HALF} + \delta) -\cos(x_{i+\HALF} + \delta)}
                     H(\pm\delta)
  \end{array}
\end{equation}
where $H()$ is the Heaviside step function.



\end{document}